\documentstyle{ioplppt}
\input epsf

\def\id{\mathop{\rm id}\nolimits}
\def\Re{\mathop{\rm Re}\nolimits}
\def\Im{\mathop{\rm Im}\nolimits}

\def\bbbr{{\rm I\kern-0.23em R}}
\def\bbbn{{\rm I\kern-0.23em N}}
\def\bbbz{{\mathchoice {\hbox{$\sf\textstyle Z\kern-0.4em Z$}}
        {\hbox{$\sf\textstyle Z\kern-0.4em Z$}}
        {\hbox{$\sf\scriptstyle Z\kern-0.3em Z$}}
        {\hbox{$\sf\scriptscriptstyle Z\kern-0.2em Z$}}}}
\def\bbbc{{\mathchoice {\setbox0=\hbox{$\displaystyle\rm C$}\hbox{\hbox
        to0pt{\kern0.4\wd0\vrule height0.95\ht0\hss}\box0}}
        {\setbox0=\hbox{$\textstyle\rm C$}\hbox{\hbox
        to0pt{\kern0.4\wd0\vrule height0.95\ht0\hss}\box0}}
        {\setbox0=\hbox{$\scriptstyle\rm C$}\hbox{\hbox
        to0pt{\kern0.4\wd0\vrule height0.95\ht0\hss}\box0}}
        {\setbox0=\hbox{$\scriptscriptstyle\rm C$}\hbox{\hbox
        to0pt{\kern0.4\wd0\vrule height0.95\ht0\hss}\box0}}}}

\newcommand{\C}{\bbbc}

\newcommand{\R}{\bbbr}

\newcommand{\proof}{\noindent{\bf Proof: }}
\newcommand{\qed}{\quad\rule{1.5mm}{1.5mm}}
\newcommand{\ds}{\displaystyle}

\newcommand{\la}{\lambda}

\newcommand{\al}{\alpha}
\newcommand{\ep}{\epsilon}
\newcommand{\ga}{\gamma}
\newcommand{\del}{\delta}

\newcommand{\sig}{\sigma}
\newcommand{\Sig}{\Sigma}

\newcommand{\Ga}{\Gamma}

\newcommand{\om}{\omega}
\newcommand{\M}{{\cal M}}

\newcommand{\Me}{\M_\ep}

\newcommand{\bfu}{{\bf u}}

\newcommand{\bfh}{{\bf H}}
\newcommand{\bfg}{{\bf G}}
\newcommand{\bfy}{{\bf Y}}

\newcommand{\bfpsi}{{\bf\Psi}}
\newcommand{\bfz}{{\bf Z}}

\newcommand{\Rk}{{\cal R}_K}

\begin{document}
\jl{8}
\title{Existence and stability of hole solutions to 
       complex Ginzburg-Landau equations}[Existence and stability
       of hole solutions]

%       \thanks{1991 Mathematics Subject Classification:
%       30B10, 30B40, 34A05, 34A26, 34A47, 34C35, 34C37, 34D15, 34E05, 
%       35K57, 35P15, 35Q51, 35Q55, 78A60}}
       
\author{Todd Kapitula\dag and Jonathan Rubin\ddag}

\address{\dag Department of Mathematics and Statistics,
    University of New Mexico, Albuquerque, NM 87131, USA,
    E-mail: {\tt kapitula@math.unm.edu}}
\address{\ddag Department of Mathematics, Ohio State University,
    Columbus, OH 43210, USA,
    E-mail: {\tt jrubin@math.ohio-state.edu}}

\begin{abstract}

We consider the existence and stability of the hole, or dark soliton,
solution to a Ginzburg-Landau perturbation of the defocusing 
nonlinear Schr\"odinger equation (NLS), and to the nearly real complex 
Ginzburg-Landau equation (CGL).
By using dynamical systems techniques, it is shown that the dark soliton 
can persist as either a regular perturbation or a singular perturbation 
of that which exists for the NLS.
When considering the stability of the soliton, a major difficulty
which must be overcome is that eigenvalues may bifurcate out of
the continuous spectrum, i.e., an {\it edge bifurcation} may occur.
Since the continuous spectrum for the NLS covers the imaginary axis,
and since for the CGL it touches the origin, such a bifurcation may
lead to an unstable wave.
An additional important consideration is that an edge bifurcation 
can happen even if there are no eigenvalues embedded in the continuous
spectrum.
Building on and refining ideas first presented in Kapitula and Sandstede
\cite{kapitula:sob98} and Kapitula \cite{kapitula:tef99}, we 
show that when the wave persists as a regular perturbation, at most three 
eigenvalues will bifurcate out of the continuous spectrum.
Furthermore, we precisely track these bifurcating eigenvalues, and thus
are able to give conditions for which the perturbed wave will be stable.
For the NLS the results are an improvement and refinement of previous work, 
while the results for the CGL are new.
The techniques presented are very general and are therefore
applicable to a much larger class of problems than those considered
here.

\end{abstract}

\ams{30B10, 30B40, 34A05, 34A26, 34A47, 34C35, 34C37, 34D15, 34E05, 
       35K57, 35P15, 35Q51, 35Q55, 78A60}
\maketitle

%\author{Todd Kapitula{ 
%       \thanks{E-mail: kapitula@math.unm.edu. Research 
%        partially supported under NSF grant DMS-9803408}}
%   \\ Department of Mathematics and Statistics
%   \\ University of New Mexico
%   \\ Albuquerque, NM 87131
%   \\ USA
%   \\
%   \\ Jonathan Rubin{
%       \thanks{E-mail: jrubin@math.ohio-state.edu. Research 
%        partially supported under NSF grant DMS-9804447}}
%   \\ Department of Mathematics
%   \\ Ohio State University
%   \\ Columbus, OH 43210
%   \\ USA}

%\date{ }
%\date{\today }

\newtheorem{theorem}{Theorem}[section]
\newtheorem{lemma}[theorem]{Lemma}
\newtheorem{hyp}[theorem]{Hypothesis}
\newtheorem{prop}[theorem]{Proposition}
\newtheorem{cor}[theorem]{Corollary}
\newtheorem{conjecture}[theorem]{Conjecture}
\newtheorem{defin}[theorem]{Definition}
\newtheorem{remark}[theorem]{Remark}
\newtheorem{ass}[theorem]{Assumption}
\newtheorem{fig}[theorem]{Figure}

%\maketitle

\renewcommand{\theequation}{\arabic{section}.\arabic{equation}}

%%%%%%%%%%%%%%%%%%
% Introduction
%%%%%%%%%%%%%%%%%%%

\section{Introduction}
\setcounter{equation}{0}

The standard model for the propagation of pulses in an ideal
defocussing nonlinear fiber without loss is the cubic nonlinear
Schr\"odinger equation (NLS)
\begin{equation}\label{eq:NLS}
i\phi_t-\frac12\phi_{xx}-\phi+|\phi|^2\phi=0,
\end{equation}
for $x\in\R$.
It supports the dark soliton solution, which is given by
\begin{equation}\label{eq:def_DS}
\Phi(x)=\tanh(x).
\end{equation}
If loss is present in the fiber, then the dark soliton will cease to exist.
Thus, at a minimum amplifiers must be used to compensate for the loss.
The effects of linear loss in the fiber as well as 
linear and nonlinear amplification of the wave along the
fiber will be incorporated into the model.
The issues to be discussed in this paper are the
persistence of the dark soliton under perturbation, and
the stability of the persisting solution relative to the PDE.
In this article, we shall concentrate on these issues for a
particular perturbation.
We emphasize, however, that the methods and ideas presented
herein are general, and they are applicable to a much larger class of problems.
Here we will consider a perturbed NLS (PNLS) which is given by
\begin{equation}\label{eq:NLS_pert}
i\phi_t-\frac12\phi_{xx}-\phi+|\phi|^2\phi=
i\ep(\frac12d_1\phi_{xx}+d_2\phi+d_3|\phi|^2\phi+d_4|\phi|^4\phi),
\end{equation}
where $\ep>0$ is small and the other parameters are real and of $O(1)$
in $\ep$.
The nonnegative parameter $d_1$ describes spectral filtering, $d_2$
describes the linear gain ($d_2>0$) or loss ($d_2<0$) due to the
fiber, and $d_3$ and $d_4$ describe the nonlinear gain or loss due
to the fiber.

A related equation is the nearly real complex Ginzburg-Landau
equation (CGL)
\begin{equation}\label{eq:GL}
\phi_t-\frac12\phi_{xx}-\phi+|\phi|^2\phi=
i\ep(\frac12d_1\phi_{xx}+d_2\phi+d_3|\phi|^2\phi+d_4|\phi|^4\phi),
\end{equation}
where again $\ep>0$ is small and the other parameters are real and
of $O(1)$.
The CGL governs the nonlinear evolution of perturbations of a simple
solution of a basic system of partial differential equations at
near critical conditions, provided that the basic system satisfies
some generic conditions (Eckhaus \cite{eckhaus:ome92}).
The CGL has been proven valid in an asymptotic sense for a large
class of systems (Collet and Eckmann \cite{collet:ttd90}, van
Harten \cite{harten:otv91}, Bollerman \etal \cite{bollerman:otj95},
Mielke and Schneider \cite{mielke:afm95}, and Schneider
\cite{schneider:eef94, schneider:gev94}).
The CGL results from an asymptotic expansion, and equation
(\ref{eq:GL}) with $d_4=0$ is only the $O(1)$ part of a more
extended equation.
The inclusion of the $d_4$ term is a means of modelling the
effect of small, nonlinear higher order corrections
(Doelman \cite{doelman:bth96}, Popp \etal \cite{popp:hsi95}, 
Stiller \etal \cite{stiller:awk95, stiller:ftd95}).

It is clear that studying the existence of steady-state
solutions to equations (\ref{eq:NLS_pert}) and (\ref{eq:GL}) amounts to
determining the solution structure for the equation
\begin{equation}\label{eq:GL_ode}
-\frac12\phi''-\phi+|\phi|^2\phi=
i\ep(\frac12d_1\phi''+d_2\phi+d_3|\phi|^2\phi+d_4|\phi|^4\phi)
\end{equation}
($'=\frac{d}{dx}$).
To do this, one can set 
\[
\phi(x)=r(x)e^{i\int_0^x\psi(s)\,ds},
\]
and then study trajectories in the
$(r,r',\psi)$ phase space.
This task has been done in a series of papers, of which
Doelman and Doelman \etal \cite{doelman:stp89, doelman:twc93, 
doelman:bth96, doelman:paq91},
Duan \etal \cite{duan:fdw95}, Holmes \cite{holmes:sst86},
Jones \etal \cite{jones:nrf90},
Kapitula and Kapitula \etal \cite{kapitula:sho95, kapitula:bba98, 
kapitula:sdo96},
Marcq \etal \cite{marcq:eso94}, and Van Saarloos \etal
\cite{saarloos:fps92} are a sample.
In Section 2 we prove the following theorem regarding the 
persistence of the wave given by (\ref{eq:def_DS}).
The result is not entirely new, as it is alluded to by
Doelman \cite{doelman:bth96}.
To determine the stability of the perturbed
waves relative to the PDEs, however, we need more detailed 
asymptotic information than
that which is provided in \cite{doelman:bth96}.

\begin{theorem}\label{thm:existence} Suppose that
\[
d_2+d_3+d_4=-\ep^2\sig^*(\ep)-\sig,
\]
where
\[
\sig^*(0)=-\frac29(d_1+d_3+\frac85d_4)^2(d_1+d_3+2d_4).
\]
Suppose that $(\ep^2\sig^*(\ep)+\sig)(d_1+d_3+2d_4)<0$.
If $\sig=0$, then the wave persists as a regular perturbation, with 
the asymptotic expansion
\[
\begin{array}{lll}
r(x)&=&\Phi(x)+O(\ep^2) \\
\vspace{.1mm} \\
\psi(x)&=&{\ds\frac{2}{3}\left((d_1+d_3+d_4)\Phi(x)
   +\frac35d_4\Phi^3(x)\right)\ep+O(\ep^3)}.
\end{array}
\]
If $\sig\neq0$, then the wave persists as a singular perturbation.
\end{theorem}

\begin{remark} When $\sig\neq0$, the radial profile of the wave
will have a ``shelf'' (\cite{burtsev:ndo97, chen:eon98, ikeda:tco95, 
ikeda:sod97}).
\end{remark}

\begin{remark} The wave $-\Phi$, which exists for $\ep=0$, persists 
under the same conditions; our analysis shows that it has the same
stability characteristics as $\Phi$ as well.  
For concreteness, we will simply refer to $\Phi$ throughout this paper. 
\end{remark}

It seems that all previous attempts to consider the stability of the wave, 
especially for the PNLS,
have ignored the 
fact that the wave persists as a singular perturbation except on
the regular perturbation manifold $d_2+d_3+d_4=-\ep^2\sig^*$;
relevant works include Burtsev \etal 
\cite{burtsev:ndo97},
Chen \etal \cite{chen:eon98}, Ikeda \etal \cite{ikeda:tco95, ikeda:sod97},
and Lega \etal \cite{lega:ths97}.
If the parameters do not lie on the regular perturbation manifold, then
it may be the case that the ``shelf'' can influence 
the stability of the wave.
One possible way of attacking this problem may be through the topological
methods first introduced by Jones \cite{jones:stw84} and 
Alexander \etal \cite{alexander:ati90}, and later used in a variety
of contexts by, for example, Bose \etal \cite{bose:sot95},
Doelman \etal \cite{doelman:saoII},
Gardner and Gardner \etal \cite{gardner:shb92, gardner:two89, gardner:stw91},
and Rubin and Rubin \etal \cite{rubin:sab, rubin:bae97}.
This issue will not be addressed in this paper and will be a topic
of future study.

Here, we suppose that the wave does persist as a regular perturbation.
Since the equations under consideration are posed on the unbounded real
line, the spectrum of the linearization about the wave contains continuous
spectrum corresponding to radiation modes.
In addition, the spectrum may contain several isolated eigenvalues of
finite multiplicity.
Because of the translation and rotation invariance of the PNLS and CGL,
zero is an eigenvalue.
It is {\it not}, however, an isolated eigenvalue.
When $\ep=0$, the continuous spectrum for the NLS covers the 
imaginary axis, while that for the CGL covers the negative real axis.
Furthermore, there are no point eigenvalues in the open right-half plane
for either equation.
For $\ep\neq0$, the origin is still contained in the continuous
spectrum. 
By choosing the parameters appropriately, one can bound the
continuous spectrum in the closed left-half plane.
To determine the stability of the wave for $\ep\neq0$, it is
thus necessary to locate the point eigenvalues.
There are standard tools available which can be used to determine
the fate of isolated eigenvalues (see, for example, Kapitula 
\cite{kapitula:tef99}).
However, it is a difficult and nonstandard problem to determine
the conditions under which eigenvalues can bifurcate out of the
continuous spectrum, i.e., conditions under which an {\it edge bifurcation}
can occur.
The primary issue of this paper is the detection of such eigenvalues.
We emphasize that an edge bifurcation may occur even if the
corresponding eigenfunctions in the unperturbed problem are not localized.

We now turn to an outline of our approach for locating eigenvalues.
In many respects it follows that presented in Kapitula \etal 
\cite{kapitula:sob98}, which deals with the stability of solitary wave
solutions for the focusing NLS.
The major tool that we use is the Evans function, $E(\la)$.
The Evans function is a complex-valued function depending on
$\la\in\C$ with the property that $E(\la)=0$ whenever $\la$ is an
isolated eigenvalue.
It is only defined a-priori away from the continuous spectrum, so it
is not immediately clear that it can be used to locate embedded
eigenvalues and detect edge bifurcations.
However, as an application of the Gap Lemma, discovered simultaneously
and independently by Kapitula \etal \cite{kapitula:sob98} and
Gardner \etal \cite{gardner:tgl98}, the Evans function can be
analytically extended across the continuous spectrum.
The analytic extension can then in theory be used to locate embedded eigenvalues
and to track them under perturbation.

In the problems considered so far, it turns out that the continuous 
spectrum corresponds
to a branch cut for the Evans function.
Furthermore, in these problems it is only at the
branch point that the Evans function has an embedded zero, so only
from there can an eigenvalue bifurcate.
For the problems under consideration both in this paper and in
Kapitula \etal \cite{kapitula:sob98}, when $\ep=0$ the edge of
the continuous spectrum is a branch point of order one, i.e.,
near the edge of the continuous spectrum we can write
$E(\la)=f(\sqrt{\la-\la_b})$, where $f(\cdot)$ is analytic and
$\la_b$ is the branch point.
In \cite{kapitula:sob98} the stability of the solitary
wave to the perturbed focusing NLS was considered.
It turned out that for a suitably scaled eigenvalue parameter that
near the branch point $\la_b=i\om$ the Evans
function could be written as
\[
E(\la,\ep)=\sqrt{\la-i\om}+A\ep,
\]
where $A\in\C$ depended upon the particular perturbation.
Thus, for that problem at most one eigenvalue could pop out of
the continuous spectrum.

To determine the location of the zeros of $E(\la)$ 
near $\la_b$ for those problems in which more than one eigenvalue can pop
out of the continuous spectrum, one would like to write the Evans function as
the series
\[
E(\ga)=\sum_{n=0}^\infty a_n\ga^n,\quad \ga^2=\la-\la_b,
\]
and then locate its zeros.
This task can be accomplished if one can derive asymptotic expressions
for the coefficients of the series.
Fortunately, by suitably modifying the ideas and methods of Kapitula 
\cite{kapitula:tef99}, which were developed for doing Taylor expansions
around isolated eigenvalues, we are able to derive such expressions.
Once the zeros of the expansion have been located, we take
those zeros that lie on the correct sheet of the appropriate 
Riemann surface and 
invert to find the eigenvalues for the system.
The interested reader should consult Section 3 for more details.

It turns out, for both the PNLS and the CGL, that when $\ep=0$ 
the Evans function has a branch point at $\la=0$ and is nonzero 
everywhere else in the closed right-half plane.
Furthermore, when $\ep=0$ the Evans function has the expansion
\[
E(\ga)=A\ga^3+O(\ga^4),
\]
where $A\in\R$ and $\ga$ is a suitably defined
function of $\la$  for $\la$ near zero (see Section 3 for details).
Thus, for the perturbed problem, there will be three zeros of the
Evans function near
$\ga=0$, and hence there will be at most three eigenvalues in this region.
By computing the lower order terms in the series, we are able
to locate these eigenvalues and assess the stability of the hole solution.
As the following theorem illustrates, for the PNLS there are
at most two eigenvalues which bifurcate out of the branch point
$\la=0$ and leave the continuous spectrum.
Furthermore, the $d_4$ term must be nonzero (specifically, negative) for the 
wave to be 
linearly stable.

\begin{figure}[htb]
\epsfysize=7.5in
%%\epsffile{d:/papers/gl_front/z_0flow.ps}
\epsffile{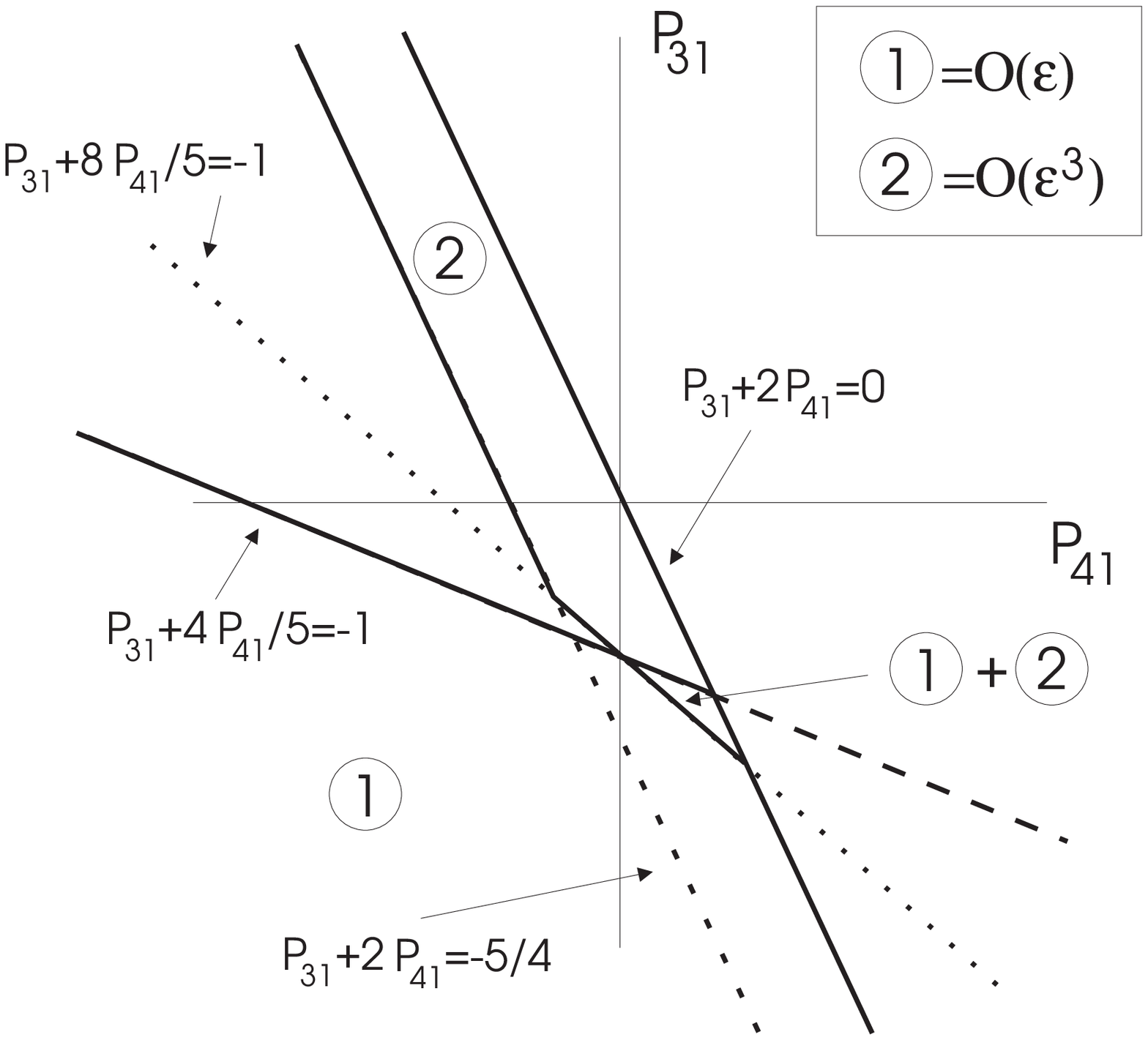}
\caption{Stability regime for NLS ($d_1>0$)
     \label{fig:nls}}
\end{figure}

\begin{theorem}\label{thm:evals_NLS} Suppose that
$d_2+d_3+d_4=-\ep^2\sig^*(\ep)$, where $\sig^*$ is given in 
Theorem \ref{thm:existence}.
Also, assume that $d_3+2d_4<0$. 
\begin{description}
\item[i)] Suppose that $d_1>0$,
and set $P_{j1}=d_j/d_1$.
If 
\[
P_{31}<-\frac45P_{41}-1,
\]
then the linearization of (\ref{eq:NLS_pert}) about the perturbed wave yields 
a positive $O(\ep)$ real eigenvalue given to leading order by 
\[
\la_1=-(d_3+2d_4)\left(\sqrt{1+\frac49\frac{(1+P_{31}+4P_{41}/5)^2}
{(P_{31}+2P_{41})^2}}-1\right)\,\ep.
\]
Furthermore, if
\[
P_{31}>-\frac85P_{41}-1,\quad P_{31}>-2P_{41}-\frac54,
\]
then there is a positive $O(\ep^3)$ real eigenvalue which is given to 
leading order by
\[
\la_2=-\frac{\tilde{\ga}}{2(P_{31}+2P_{41})}\ep^3,
\]
where
\[
\tilde{\ga}=\frac49 d_1^3(1+P_{31}+\frac85P_{41})^2
  (\frac54+P_{31}+2P_{41}).
\]
Otherwise, the wave is linearly stable, as no other eigenvalues bifurcate
from the continuous spectrum (see Figure \ref{fig:nls}).
\item[ii)] If $d_1=0$, then the wave is linearly stable as a solution of
(\ref{eq:NLS_pert}) if $5d_3+4d_4<0$; otherwise,
there is an $O(\ep)$ eigenvalue which is given to leading order by 
\[
\la_1=-(d_3+2d_4)\left(\sqrt{1+\frac49\frac{(d_3+4d_4/5)^2}
{(d_3+2d_4)^2}}-1\right)\,\ep.
\]
\end{description} 
\end{theorem}

\begin{remark} The condition that $d_1\ge0$ and $d_3+2d_4<0$ ensures
that the continuous spectrum is contained in the closed left-half plane
for $\ep>0$ and small.
\end{remark}

\begin{remark} If $d_4=0$ the wave is linearly unstable, with an
$O(\ep)$ eigenvalue if $P_{31}<-1$ and an $O(\ep^3)$ eigenvalue if
$-1<P_{31}<0$.
Furthermore, the wave is linearly unstable if $d_4>0$.
\end{remark}

Before we discuss the stability of the wave for the CGL, a few comments
are in order.
There have been many recent efforts to determine the stability of the
dark soliton for the perturbed NLS by using an adiabatic approach
(\cite{burtsev:ndo97, chen:eon98, ikeda:tco95, ikeda:sod97, lega:ths97}).
With the adiabatic approach the wave is predicted to be stable if both
$d_3+2d_4<0$ and  
$d_1+d_3+6d_4/5>0$ hold.
If $d_4=0$, then this approach is consistent with the result of
Theorem \ref{thm:evals_NLS} in that it correctly determines the
stability of the wave up to $O(\ep)$.
However, it does not predict the existence of the $O(\ep^3)$ instability; 
this is not surprising, as the adiabatic approach is only meant
to understand the dynamics on a time scale of $O(1/\ep)$.
If $d_4\neq0$, then the analysis contradicts the results presented in this 
paper, even at the $O(\ep)$ level.
This contradiction implies that the original adiabatic ansatz for the
slow-time variation of the wave is incorrect (see Section 5.5 for
more details).
In some way the parameter $d_4$ has the same effect on
the stability analysis for the perturbed wave as it has on  
the solution structure for the steady-state problem,
i.e., it breaks some kind of ``hidden symmetry'' (see
Doelman \cite{doelman:bth96}).
This topic would be an interesting avenue for further research.

When considering the stability of the wave to the CGL, the primary
difficulty is that the resulting Evans function is not as easy
to factor as that associated with the PNLS.
As such, for general parameter values the location of bifurcating 
eigenvalues cannot be put into an easily readable form.
However, one can determine for which ranges in
the parameter space there will be eigenvalues with positive
real part; as with the PNLS, it turns out that at most two eigenvalues bifurcate 
from the continuous spectrum.
As it can be seen from the following theorem, a primary difference
between the PNLS and the CGL when considering the stability of the 
hole solution is the order of the eigenvalues.
In general, the instability will grow much more slowly for the
CGL than for the PNLS.

\begin{figure}[htb]
\epsfysize=7.5in
%%\epsffile{d:/papers/gl_front/z_0flow.ps}
\epsffile{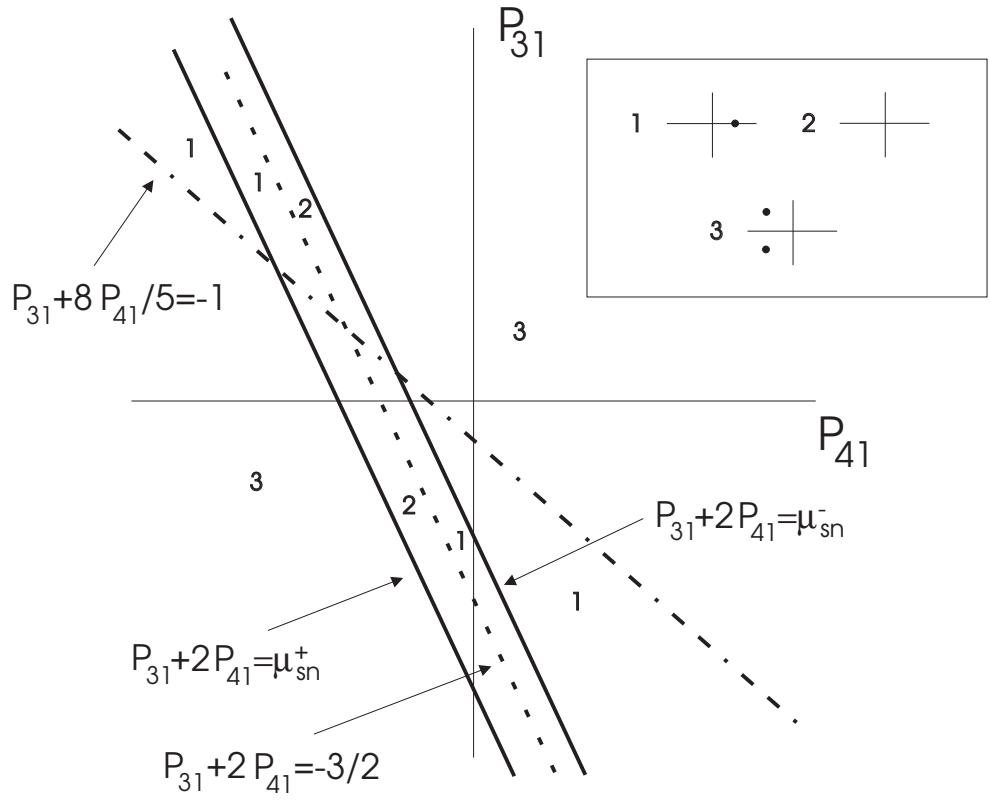}
\caption{Stability regime for CGL ($d_1>0$)
     \label{fig:cgl2}}
\end{figure}

\begin{theorem}\label{thm:evals_GL} Suppose that
$d_2+d_3+d_4=-\ep^2\sig^*(\ep)$, where $\sig^*$ is given in 
Theorem \ref{thm:existence}.
Set
\[
\mu_{sn}^\pm=\frac32\frac{\pm\al-2/3}{1\mp\al},\quad
  \al^2=\frac{\sqrt{125}+11}2
\]
($\mu_{sn}^+=-1.716,\,\mu_{sn}^-=-1.385$).

\begin{description}
\item[i)] Suppose that $d_1\neq0$, and set $P_{j1}=d_j/d_1$.
If
\[
(\frac32 +P_{31}+2P_{41})(1+P_{31}+\frac85P_{41})<0,
\]
then there is one positive real $O(\ep^4)$ eigenvalue 
for the linearized problem, and the wave is linearly unstable.
If
\[
d_1(1+P_{31}+\frac85P_{41})>0,\quad d_1(\mu_{sn}^- +P_{31}+2P_{41})>0
\]
or 
\[
d_1(1+P_{31}+\frac85P_{41})<0,\quad d_1(\mu_{sn}^+ +P_{31}+2P_{41})<0,
\]
then there is a complex pair of $O(\ep^4)$ eigenvalues with negative
real part.
Otherwise, no eigenvalues bifurcate from the continuous spectrum
(see Figure \ref{fig:cgl2}).
In either case, if 
\[
(\frac32 +P_{31}+2P_{41})(1+P_{31}+\frac85P_{41})>0,
\]
then the wave is linearly stable.

\item[ii)] Suppose that $d_1=0$ and set
\[
a=(d_3+2d_4)(d_3+\frac85d_4).
\]
If $a>0$, then the zeros of the Evans function inside the curve $K$ 
are given by
\[
\la_{2,3}=(-0.595\pm0.255i)\,a^2\ep^4,
\]
and the wave is linearly stable as a solution of (\ref{eq:GL}).
If $a<0$, then the zero of the Evans function inside $K$ is given by
\[
\la_1=1.191\,a^2\ep^4,
\]
and the wave is linearly unstable.
\end{description}
\end{theorem}

\begin{remark} The continuous spectrum remains in the closed left-half
plane for all values of $d_1,\dots,d_4$ as long as $\ep>0$ is sufficiently
small.
\end{remark}

\begin{remark} The sign of the parameter $a$ corresponds to
the manner in which the wave is constructed in the $(r,r',\psi)$
phase space.
The interested reader should consult Section 2 for more details.
\end{remark}

\begin{remark} If $d_1\neq0$, it may be the case that there is
a complex pair of eigenvalues with negative real part.
The interested reader should consult Lemma \ref{lem:evans_roots_d1not0}
for the details.
\end{remark}

The remainder of this paper is organized in the following manner.
In Section 2 the conditions for the persistence of the wave are
derived through the use of dynamical systems techniques.
In Section 3 we derive the expressions which allow us to compute Taylor
expansions at the branch point of the Evans function.
This section is relatively self-contained and can be skipped on a
first reading.
In Sections 4 and 5 we calculate the Taylor expansion for the Evans
function for the CGL and the PNLS, respectively.
Theorem \ref{thm:evals_GL} follows from Lemmas \ref{lem:evans_roots_d1=0} 
and \ref{lem:evans_roots_d1not0}.
Theorem \ref{thm:evals_NLS} follows from Lemma \ref{lem:evans_roots_NLS}.
Section 5 concludes with a brief discussion comparing the approach
of this paper with the previous adiabatic approaches.

\begin{remark} Recently, Li and Promislow \cite{li:tmo} independently
and simultaneously used some of the ideas present in this paper to
study the stability of waves to the equations describing pulse
propagation in linearly birefringent, lossless fibers.
\end{remark}

\section{Existence and persistence}
\setcounter{equation}{0}

The steady-state problem for both the PNLS and the CGL is given by
\begin{equation}\label{eq:pcgl}
-\frac12\phi''-\phi+|\phi|^2\phi=i\ep 
   (\frac12d_1\phi''+d_2\phi+d_3|\phi|^2\phi+d_4|\phi|^4\phi)
\end{equation}
($'=\frac{d}{dx}$).
For existence of the hole solution, which is given by
\begin{equation}\label{eq:def_Phi}
\Phi(x)=\tanh x
\end{equation}
when $\ep=0$, we will want to consider the problem in polar coordinates.
Set
\begin{equation}\label{eq:polar_rep}
\phi(x)=r(x)e^{i\int_0^x\psi(s)\,ds}
\end{equation}
to obtain (after dropping higher order terms that do not affect subsequent
calculations) the three-dimensional system of ODEs
\begin{equation}\label{eq:pcgl_ode}
\begin{array}{rll}
r'&=&s \\
s'&=&-2r(1-r^2)+r\psi^2 \\
&&\quad-2\ep^2d_1r(d_2-d_1+(d_1+d_3)r^2+d_4r^4) \\
\psi'&=&-2\frac{s}{r}\psi-2\ep(d_2-d_1+(d_1+d_3)r^2+d_4r^4).
\end{array}
\end{equation}

For the system (\ref{eq:pcgl_ode}) there exist are two critical 
manifolds $\M_\ep^\pm$, which when $\ep=0$ are given by
\begin{equation}\label{eq:def_Me}
\M^\pm_0=\{(r,s,\psi)\,:\,r=\pm\sqrt{1-\psi^2/2},\,\psi^2<2/3\};
\end{equation}
we restrict to $\psi^2<2/3$ in (\ref{eq:def_Me}) so that the manifolds
$\M_\ep^\pm$ are normally hyperbolic.
Each critical manifold of (\ref{eq:pcgl_ode}) has a two-dimensional 
unstable manifold, 
$W^u(\M^\pm_\ep)$, and a two-dimensional stable manifold, $W^s(\M^\pm_\ep)$,
which are smooth perturbations of the center-stable and center-unstable
manifolds which exist when $\ep=0$ \cite{fenichel:pas73, jones:gsp95}.
As it will be seen, it can be shown that 
$W^u(\M^-_\ep)\cap W^s(\M^+_\ep)\neq\emptyset$, and, by the symmetry
$(r,s,\psi,x)\to(r,-s,-\psi,-x),\,
W^u(\M^+_\ep)\cap W^s(\M^-_\ep)\neq\emptyset$, 
both for $0\le\ep<\ep_0$ for some $\ep_0>0$.
These relationships are clearly satisfied when $\ep=0$, as evidenced
by the existence of the waves $\pm\Phi$.
Assuming that the relevant manifolds intersect, the wave $\Phi$ will
persist as long as the parameters are chosen so that critical points
exist on $\M^\pm_\ep$ (also see Doelman \cite{doelman:stp89, doelman:twc93}).
Depending how the parameters are chosen, there will be zero, two, or
four critical points on $\M^\pm_\ep$ (counting multiplicities).
The condition $\psi^2<2/3$ implies that the critical points on $\M^\pm_\ep$
correspond to stable periodic solutions to (\ref{eq:pcgl})
\cite{kapitula:ons94, kapitula:eas96}.

To prove the existence of multiple orbits bifurcating from
the original heteroclinic cycle with the constraint that the orbits remain
within an small tube of the original cycle, it will be useful
to set 
\begin{equation}\label{eq:def_d2*}
d_2+d_3+d_4=-(\ep^2\sig^*+\sig),
\end{equation}
where $\sig^*(\ep)$ is such that
\begin{equation}\label{eq:def_sig*}
\sig^*(0)=-\frac29(d_1+d_3+\frac85d_4)^2(d_1+d_3+2d_4),
\end{equation}
as in the statement of Theorem \ref{thm:existence}.
It will henceforth be assumed that the parameter $\sig$, while small, is
independent of $\ep$.

\begin{remark}
Equation (\ref{eq:def_d2*}) is not a
parameter restriction for the CGL, as it can always be achieved by
going into an appropriate rotating reference frame.
However, it is a restriction for the PNLS, and determines a balance
between the linear loss and nonlinear gain.
\end{remark}

Substituting relation (\ref{eq:def_d2*}) into the ODE (\ref{eq:pcgl_ode}) 
yields 
\begin{equation}\label{eq:pcgl_ode_new}
\begin{array}{rll}
r'&=&s \\
s'&=&-2r(1-r^2)+r\psi^2 \\
&&\quad
      +2\ep^2d_1r[(d_1+d_3)(1-r^2)+d_4(1-r^4)+\ep^2\sig^*+\sig] \\
\psi'&=&-2\frac{s}{r}\psi+2\ep[(d_1+d_3)(1-r^2)+d_4(1-r^4)+\ep^2\sig^*+\sig].
\end{array}
\end{equation}
Since the lowest order at which
$\sig$ appears in (\ref{eq:pcgl_ode_new}) is at $O(\ep)$ in the $\psi$-equation, 
the effect of $\sig$ on perturbation calculations will only
be felt at $O(\ep+\sig)\ep$,  
except in terms of the location of critical points on $\Me$, which is discussed
below.
Hence, for many of the perturbation calculations that follow, the
role of $\sig$ can be ignored.

The following two propositions detail the relevant behavior on 
$\M^\pm_\ep$.
The proofs can be found in Kapitula \cite{kapitula:bba98} and hence 
are omitted.

\begin{prop}\label{prop:crit_pts} Suppose that $d_2+d_3+d_4=-(\ep^2\sig^*
+\sig)$
and that
\[
(\ep^2\sig^*+\sig) (d_1+d_3+2d_4)<0.
\]
Then a pair of critical points on $\M^+_\ep$ [$\M^-_\ep$] are given by 
$(r^*_+,0,\pm\psi^*)$ [$(r^*_-,0,\pm\psi^*)$],
where
\[
\begin{array}{lll}
{\ds r^*_\pm}&=&{\pm\left(\ds 1
   +\frac12\frac{\ep^2\sig^*+\sig}{d_1+d_3+2d_4}\right)} \\
\vspace{.1mm} \\
\psi^*&=&{\ds\sqrt{-2\frac{\ep^2\sig^*+\sig}{d_1+d_3+2d_4}}}.
\end{array}
\]
\end{prop}

\begin{prop}\label{prop:Mep} When $0\le\ep\ll 1$, the manifolds $\M^\pm_\ep$
intersect the $r$-axis.
Further, there exists $\delta$, with  $1\gg\del>0$, such that for 
$-(\psi^*+\del)<\psi<\psi^* +\del$ the flow on $\M^\pm_\ep$ is given by
\[
\psi'=\ep((d_1+d_3+2d_4)\psi^2+2\ep^2\sig^*+2\sig).
\]
\end{prop}

Proposition \ref{prop:crit_pts} gives a condition for the existence
of critical points on $\M^+_\ep$.
It remains to show that $W^u(\M^-_\ep)\cap W^s(\M^+_\ep)\neq\emptyset$ 
for small $\ep\neq0$.
Let $\Sig_o^p=\{(r,s,\psi):r=\psi=0\}$. 
The hole solution belongs to $\Sigma_o^p$ at $x=0$, with $s(0)\neq0$.
When $\ep=0$, the manifold $W^s(\M^+_\ep)$
intersects the curve $\Sig_o^p$ transversely in $(r,s,\psi)$-space,
since $W^s(\M^+_0)$ is transverse to the invariant $\{\psi=0\}$ plane.
Thus, the intersection will persist for $\ep\neq0$ sufficiently small.
Due to invariance under $(r,s,\psi,x)\to(-r,s,-\psi,-x)$ and the
fact that $s(0)\neq0$ along the $\ep=0$ solution, it can
then be concluded that not only does $W^u(\M^-_\ep)$ also intersect
$\Sig_o^p$ transversely, but 
$W^u(\M^-_\ep)\cap W^s(\M^+_\ep)\neq\emptyset$ as well.
Hence, the hole solution will persist for
$\ep\neq0$ and small.
The result is not new (for example, see Doelman \cite{doelman:stp89}).
To determine the stability of the wave, however, more information about
the wave must be known than has previously been given.

In the remainder of this section, we finish the proof of Theorem 
(\ref{thm:existence}) 
by showing that for $\sig=0$ the perturbed wave arises as a regular 
perturbation, and then compute its asymptotics. 
We conclude with a discussion of how the nature of the intersection
that yields the wave differs in various parameter regimes; this is where
Proposition \ref{prop:Mep} is useful. 

Let an underlying hole solution be denoted by $(R,S,\Psi)$.
When evaluated at $\ep=\sig=0$, the variational equations associated with 
(\ref{eq:pcgl_ode_new}) are given by
\begin{equation}\label{eq:variation}
\begin{array}{lll}
\del r'&=&\del s \\
\del s'&=&-2(1-3R^2-\Psi^2/2)\,\del r+2R\Psi\,\del\psi \\
\del\psi'&=&2R'\Psi/R^2\,\del r-2\Psi/R\,\del s
   -2R'/R\,\del\psi \\
   &{}&\quad\quad+2[(d_1+d_3)(1-R^2)+d_4(1-R^4)]\,
      \del\ep \\
\del\ep'&=&0  \\
\del\sig'&=&0.
\end{array}
\end{equation}
Since the solution belongs to $\Sig_o^p$ at $x=0$ even for $\ep\neq0$, 
it is of interest to determine the location of the curve $\Sig_o^p$ as
the flow carries it up to the slow manifold $\M^+_\ep$.
Specifically, we wish to determine the $\psi$-coordinates of the 
points of $\Sig_o^p$ as they approach $\M^+_\ep$.
Using the fact that the $\psi$-coordinate of $\Sig_o^p$ is identically
zero when $\ep=0$, by doing a Taylor expansion we can write
that $\psi=\psi_\ep \ep+O(\ep^2)$.
From evaluation of the variational equations over the $\ep=0$ 
hole solution $\Phi$, we find
that $\psi_\ep$ satisfies the initial value problem
\begin{equation}\label{eq:sigo_flow}
\begin{array}{l}
{\ds(\Phi^2\psi_\ep)'=
    2[(d_1+d_3)(1-\Phi^2)+d_4(1-\Phi^4)]\Phi^2} \\
{\ds(\Phi^2\psi_\ep)(0)=0}.
\end{array}
\end{equation}
Upon integrating, it is seen that  
\begin{equation}\label{eq:psi_ep}
\psi_\ep(x)=\frac{2}{3}\left((d_1+d_3+d_4)\Phi(x)
   +\frac35d_4\Phi^3(x)\right).
\end{equation}

Let $0<\nu\ll 1$ be given, and let $T_\nu>0$ be such that
$1-\Phi(T_\nu)=\nu$.
That is, $T_\nu$ denotes a time when the curve $\Sig_o^p$ is within
$O(\nu)$ of the slow manifold $\M^+_\ep$.
Upon evaluating the expression for $\psi_\ep$ at $T_\nu$, it is seen that
\begin{equation}\label{eq:psi_ep_limit}
\psi_\ep(T_\nu)=\frac{2}3(d_1+d_3+\frac85d_4)+O(\nu).
\end{equation}
The following proposition has now been proved.

\begin{prop}\label{prop:Sigop} At the time $T_\nu$ such that 
$1-\Phi(T_\nu)=\nu$,
the image of the curve $\Sig_o^p$ under the flow is within an $O(\nu)$ distance
of the slow manifold
$\M^+_\ep$, and the $\psi$-coordinates 
of points on the image of $\Sig_o^p$ are given by 
\[
\psi=[\frac{2}3(d_1+d_3+\frac85d_4)+O(\nu)]\ep+O(\ep+\sig)\ep,
\]
where $0<\ep,\nu\ll 1$.
\end{prop}

First suppose that $\sig=0$.
As a consequence of the manner in which $\sig^*$ has been chosen, an application
of Propositions \ref{prop:crit_pts} and \ref{prop:Sigop} yields that
the wave will persist as a regular perturbation.
This is due to the fact that the critical points on $\M^+_\ep$ match the
expression given in Proposition \ref{prop:Sigop}.
The following lemma gives the necessary asymptotics for the perturbed
wave.
The proof is a standard application of perturbation theory, and hence
will be left to the interested reader.

\begin{lemma}\label{lem:reg_pert} Suppose that $\sig=0$.
The perturbed wave then arises as a regular perturbation and satisfies
\[
\begin{array}{lll}
r&=&\Phi+r_{\ep\ep}\ep^2/2+O(\ep^3) \\
\psi&=&\psi_\ep \ep +O(\ep^3),
\end{array}
\]
where
\[
\psi_\ep(x)=\frac{2}{3}\left((d_1+d_3+d_4)\Phi(x)
   +\frac35d_4\Phi^3(x)\right)
\]
and 
\[
\begin{array}{lll}
r_{\ep\ep}(x)&=&{\ds\frac1{225}[-5(10(d_1+d_3)^2
      +40(d_1+d_3)d_4+39d_4^2)\Phi(x)}\\
\vspace{.1mm} \\
&&\quad +8d_4(5(d_1+d_3)+8d_4)\Phi^3(x)
      +3d_4^2\Phi^5(x) \\
\vspace{.1mm} \\
&&\quad +12d_4(5(d_1+d_3)+8d_4)x\Phi'(x)] \\
\vspace{.1mm} \\
&&\quad\quad {\ds+\frac13 d_1[2d_4\Phi(x)-3(d_1+d_3+2d_4)x\Phi'(x)]\Phi'(x)}.
\end{array}
\]
\end{lemma}

\begin{remark} Note that
\[
\lim_{x\to\pm\infty}(2r_{\ep\ep}\pm\psi^2_\ep)(x)=0.
\]
This fact will be important in later calculations which deal with
improper integrals.
\end{remark}

\begin{figure}[htb]
\epsfysize=7.5in
%%\epsffile{d:/papers/gl_front/z_0flow.ps}
\epsffile{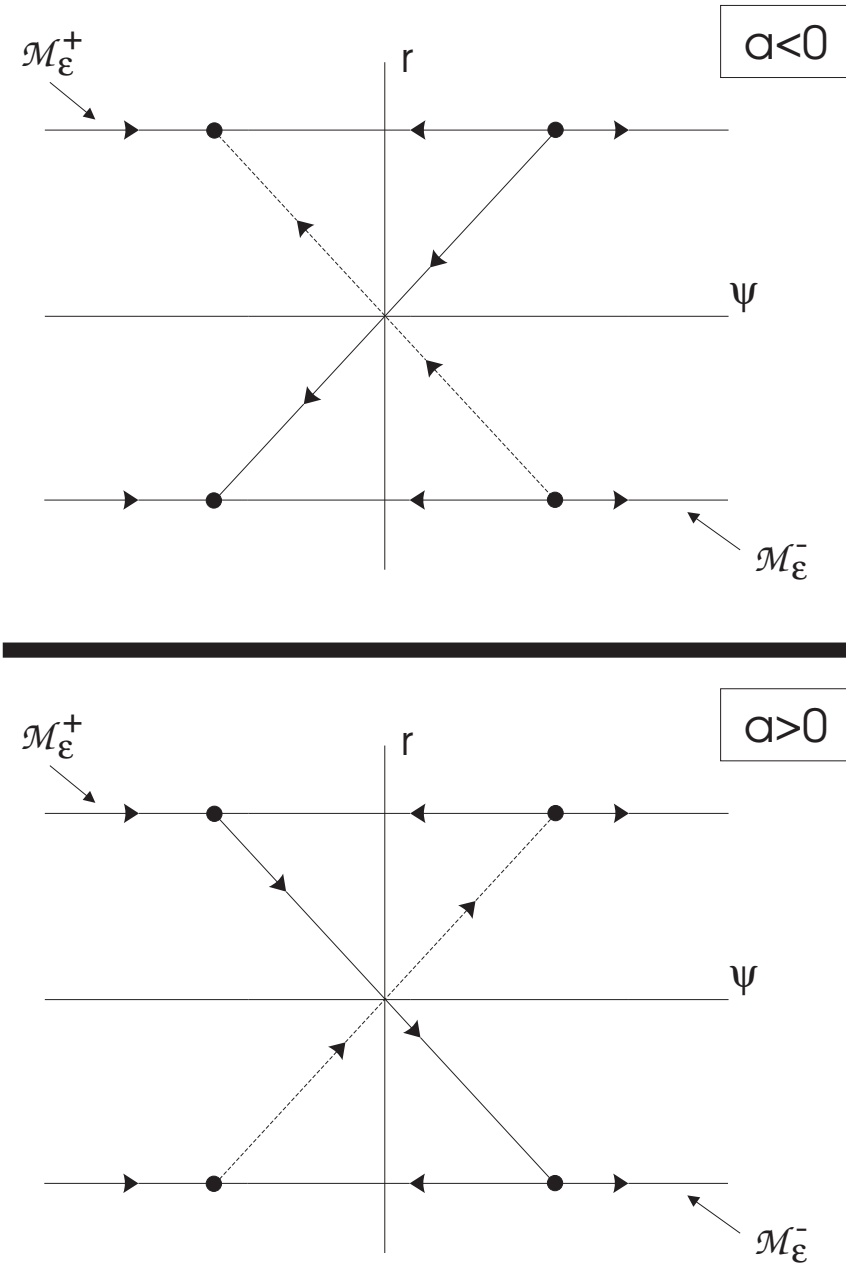}
\caption{Projected flow onto $\{s=0\}\, (a=(d_1+d_3+2d_4)(d_1+d_3+8d_4/5))$
     \label{fig:projflow}}
\end{figure}

For the rest of this paper, set
\begin{equation}\label{eq:def_psi+}
\psi_\ep^+=\lim_{x\to+\infty}\psi_\ep(x).
\end{equation}
Note that by symmetry, $\lim_{x\to-\infty}\psi_\ep(x)=-\psi_\ep^+$.
Upon doing a linear stability analysis of the critical points on 
$\M^\pm_\ep$, one notices the following facts.
If 
\begin{equation}\label{eq:regpert}
(d_1+d_3+\frac85d_4)(d_1+d_3+2d_4)<0,
\end{equation}
then the wave will be realized as the intersection of a two-dimensional unstable
manifold with a two-dimensional stable manifold in the three-dimensional 
phase space.
Alternately, if
\begin{equation}\label{eq:regpert2}
(d_1+d_3+\frac85d_4)(d_1+d_3+2d_4)>0,
\end{equation}
then the wave is realized as the intersection of a one-dimensional unstable
manifold with a one-dimensional stable manifold in the three-dimensional 
phase space.
In other words, if equation (\ref{eq:regpert}) holds, then
the trajectory out of the curve $\Sig_o^p$ intersects the
strong stable manifold of the point $(r_+^*,0,\ep\psi_\ep^+)$; furthermore,
the critical point is an attractor on the manifold $\M^+_\ep$. 
This is indicated by Proposition \ref{prop:Mep}, which gives the flow on
$\M^+_\ep$ for $|\psi|\ll 1$, and by Proposition \ref{prop:Sigop}.
If the parameters satisfy equation (\ref{eq:regpert2}), then the 
critical point is a repellor on the manifold $\M^+_\ep$
(see Figure \ref{fig:projflow}).
As we show in Sections 4 and 5, this structure plays a role
when discussing the stability of the wave.

Now suppose that $\sig\neq0$.
In this case, the wave arises as a result of
a singular perturbation, since 
$\psi_\ep(T_\nu)\neq\pm\psi^*$ at leading order in $\nu$.
If $\sig\sig^*>0$, then the resulting wave can be thought of as 
a concatenation of the solution $\Phi$ with
solutions tracking along close to the slow manifolds $\M^\pm_\ep$.
The radial profile of the solution will have a ``shelf'' at
the point at which it approaches $\M^\pm_\ep$ (see \cite{burtsev:ndo97,
chen:eon98, ikeda:tco95, ikeda:sod97} for a
discussion of the shelf in the context of the NLS and nonlinear 
optics).
Furthermore, the perturbed wave will stay within an $O(\ep)$ tube of 
the original ($\ep=0$) wave $\Phi$.
Now suppose that $\sig\sig^*<0$.
If equation (\ref{eq:regpert}) holds, then the wave will stay within
an $O(\ep)$ tube of $\Phi$.
If (\ref{eq:regpert2}) holds, however, then the wave will travel along
$\M^\pm_\ep$ to a critical point (if it exists) outside this tube.

\section{Derivatives at branch points}
\setcounter{equation}{0}

Consider the linear operator
\begin{equation}\label{eq:def_L}
L=B\partial_x^2+P(x)\partial_x+N(x),
\end{equation}
where $B$ is an invertible $n\times n$ matrix whose eigenvalues have
nonnegative real part,  and
$P(x)$ and $N(x)$ are smooth $n\times n$ matrices satisfying
\[
\lim_{x\to\pm\infty}P(x)=P_\pm,\quad
\lim_{x\to\pm\infty}N(x)=N_\pm,
\]
with the approach being exponentially fast.
Upon setting $\bfy=[u,u']^T$, where $'=d/dx$, the eigenvalue equation
$Lu=\la u$ can be rewritten as the first-order system 
\begin{equation}\label{eq:lin_eval}
\bfy'=M(\la,x)\bfy,
\end{equation}
 with
\[
M(\la,x)=\left[\begin{array}{cc} 0 & \id \\
   -B^{-1}(N(x)-\la\id) & -B^{-1}P(x) \end{array}\right].
\]

In this section, we define an Evans function for the operator $L$.
We do this  under
assumptions which imply that at least one of the matrices 
$M_\pm(\la) := \lim_{x\to\pm\infty}M(\la,x)$ 
has a pair of eigenvalues that produce a branch point for the Evans function
at a fixed value of $\la$.
In this context, we
develop a technique for differentiating the Evans function at this branch
point. 
This method then allows us, in Sections 4 and 5, to derive perturbation 
expansions on a Riemann surface
for particular Evans
functions around branch points.
These expansions are crucial in locating eigenvalues
for the corresponding linear operators. 

%%%%%%%%%%%%%%%%%%%%%%%%%%%%%%%%%%%%%%%%%%%%%%%%%%%%%%%%%%%%%%%%%%%%
\subsection{General assumptions and definition of Evans function} 

Consider the linear eigenvalue problem (\ref{eq:lin_eval}) where 
$M(\la,x)\in\C^{2n\times 2n}$ is smooth in $x$ for each fixed $\la$
and analytic in $\la$ for each fixed $x$. 
The following assumptions will be made on $M(\la,x)$. 

\begin{ass}\label{ass:M} The matrix $M(\la,x)$ satisfies:
\begin{itemize}
\item $\lim_{x\to\pm\infty}M(\la,x)=M_\pm(\la)$, with an
  exponentially fast approach 
\item If $\Re\la>0$, then $M_\pm(\la)$ has $n$ eigenvalues with positive
  real part and $n$ eigenvalues with negative real part
\item A pair of eigenvalues for $M_\pm(\la)$ are $\pm\sqrt{b(\la)}$, where
  $b(\la)$ is analytic at $\la=0$ with $b(0)=0$ and $b'(0)\neq0$, while
  the other $2n-2$ eigenvalues are analytic at $\la=0$ with nonzero real
  parts 
\item When put into Jordan canonical form, $M_\pm(0)$ has the block
  $\left[\begin{array}{cc}0&1\\0&0\end{array}\right]$.
\end{itemize}
\end{ass}

The second of these assumptions is not necessary, but it holds for
the applications of interest and we make it to simplify notation. 
The third and fourth assumptions imply that a pair of eigenvalues of 
$M_\pm(\la)$
form a branch point of the Evans function at $\la =0$.
Later in this section we will slightly relax the third
and fourth assumptions such that this holds for only one of the matrices 
$M_\pm(\la)$ (see Remark \ref{rem:weak_ass}). 
Taken together, the statements in
Assumption \ref{ass:M}  
imply that 
if $M(\la,x)$ is derived from the first order system representation of
a linear operator $L$, then
$\{0\}\in\sig_c(L)$ and is on the {\it edge} of the continuous
spectrum (see also \cite{kapitula:ani98,kapitula:sob98}).
Finally, we note that while it will not be done here, it may be possible
to extend the theory to the case where $M_\pm(0)$ have several Jordan blocks
of the type given above.
This could be useful when discussing the stability of waves satisfying  
viscous conservation laws (\cite{gardner:tgl98}).

We now construct the Evans function following the ideas presented
in \cite{alexander:ati90}.
If $\la$ is not in the continuous spectrum, then the matrices $M_\pm(\la)$ 
have no eigenvalues with zero real part.
If each has $n$ eigenvalues with positive real part and
$n$ with negative real part, then it 
is possible to define solutions $\bfy_i(\la,x)$ to equation
(\ref{eq:lin_eval}) which are analytic in $\la$ such that for
$i=1,\dots,n$
\[
\lim_{x\to-\infty}|\bfy_i(\la,x)|=0,\quad
  \bfy_1(\la,0)\wedge\cdots\wedge\bfy_n(\la,0)\neq 0,
\]
and for $i=n+1,\dots,2n$
\[
\lim_{x\to+\infty}|\bfy_i(\la,x)|=0,\quad
  \bfy_{n+1}(\la,0)\wedge\cdots\wedge\bfy_{2n}(\la,0)\neq 0.
\]
Following Alexander \etal \cite{alexander:ati90}, the Evans function is
given by
\[
E(\la)=\bfy_1(\la,0)\wedge\cdots\wedge\bfy_{2n}(\la,0).
\]
If $E(\la_0)=0$, then there exists a solution to (\ref{eq:lin_eval}) 
which decays exponentially fast as $|x|\to\infty$, and hence $\la_0$
is an eigenvalue for $L$.

If $\la$ is in the continuous spectrum, then at least one of the matrices
$M_\pm(\la)$ has an eigenvalue with zero real part, and the above 
construction breaks down.
Recently, Kapitula and Sandstede \cite{kapitula:sob98} and 
Gardner and Zumbrun \cite{gardner:tgl98} concurrently and independently
showed that the Evans function can be analytically extended into
the essential spectrum via the {\it Gap Lemma}.
The analyticity of the 
extension fails precisely when Assumption \ref{ass:M} holds,
as in this case the Evans function has a branch point.

In many applications, one of which was considered in \cite{kapitula:sob98},
the branch point is located on the imaginary axis.
Thus, under a perturbation of the wave, it is possible for eigenvalues
to move out of the branch point and into the right-half of the complex
plane, leading to an instability.
In other words, an {\it edge bifurcation} may occur \cite{kapitula:ani98}.
To locate any such bifurcating eigenvalues, our strategy is to
do a Taylor expansion for the Evans function in the vicinity of the branch 
point and then to locate the zeros of the resulting polynomial;
to expand appropriately, we must account for the presence of the
branch point (\cite{markushevich:tof85}).
In particular, if a point $\la_0$ is a branch point of order $k-1$
for the Evans function, then by setting $\ga=(\la-\la_0)^{1/k}$ one
obtains an expansion around the branch point of the form
\begin{equation}\label{eq:Evans_series}
E(\ga)=\sum_{n=0}^\infty a_n\ga^n.
\end{equation}
One can then find the zeros for $E(\ga)$ and use the inversion
relation $\la=\la_0+\ga^k$ to find the zeros for $E(\la)$.
The inversion must be done very carefully, however, as  
the zeros of the series (\ref{eq:Evans_series}) do not necessarily 
all correspond to eigenvalues
for the linearized problem (\ref{eq:lin_eval}). 

Let $K\subset\C$ be a simple closed curve which encircles the branch point
$\la_0$, such that no zeros of the Evans function belong to $K$ itself.
Furthermore, let $K$ be such that it encloses all the possible zeros
of $E(\la)$ which are contained in the right-half plane.
The existence of such a curve is guaranteed by a result in
Alexander \etal \cite{alexander:ati90}.
To be able to write the Evans function as the infinite series
given in equation (\ref{eq:Evans_series}), one must be able to
define the Evans function on a $k$-sheeted Riemann surface $\Rk$.
The surface $\Rk$ is constructed in the following manner
(\cite{markushevich:tof85}, \cite{silverman:ica72}).
Let $K_0, K_1,\dots,K_{k-1}$ be copies of $\overline{K}$ cut along the
nonpositive real axis.
Let $\del_j^\pm$ denote the upper and lower edges of the nonpositive
real axis regarded as the boundary of $K_j$, and let
\[
(\la-\la_0)^{1/k}=|\la-\la_0|^{1/k}\exp[i(\mbox{arg}\,\la-\la_0+2j\pi)/k]
\]
on $K_j$.
Now paste $\del_0^-$ to $\del_1^+,\,\del_1^-$ to $\del_2^+,\,\dots,\,
\del_{k-2}^-$ to $\del_{k-1}^+$, and finally
$\del_{k-1}^-$ to $\del_0^+$.
The result is a $k$-sheeted Riemann surface $\Rk$, with 
the sheets coming together at the branch point $\la=\la_0$.
The Gap Lemma (\cite{gardner:tgl98}, \cite{kapitula:sob98})
implies that the 
function $E(\la)$ extends analytically to the surface $\Rk$, and hence the
series is valid.
For the zeros of the series (\ref{eq:Evans_series}) to 
correspond to eigenvalues, they must lie on the correct sheet of
the Riemann surface.
In particular, they must satisfy
\begin{equation}\label{eq:zero_valid}
-\frac{\pi}{k}<\mbox{arg}\,\ga<\frac{\pi}{k},
\end{equation}
so that they are located on the sheet $K_0$.
Zeros of the series on other sheets correspond to the existence of 
solutions of 
(\ref{eq:lin_eval}) that are not eigenfunctions. 

Under Assumption \ref{ass:M},
the Evans function will be defined on a 2-sheeted Riemann
surface.
To take into account the fact that a pair of eigenvalues of $M_\pm(\la)$
has a branch point at $\la=0$, set
\begin{equation}\label{eq:def_ga}
\ga^2=b(\la).
\end{equation}
By the assumptions on the matrices $M_\pm(\la)$, for $\Re\la\ge0$ there
exist solutions $\bfy_{f,i}^\pm(\la,x),\,i=1,\dots,n-1$, such that
$|\bfy_{f,i}^\pm(\la,x)|\to 0$ exponentially fast as $x\to\pm\infty$.
From the third assumption and equation (\ref{eq:def_ga}), there also
exist solutions $\bfy_s^\pm(\ga,x)$ which satisfy
\begin{equation}\label{eq:def_ys}
\lim_{x\to\pm\infty}\bfy_s^\pm(\ga,x)e^{\pm\ga x}=v_s^\pm(\ga).
\end{equation}
The vectors $v_s^\pm(\ga)$ are analytic in $\ga$ and satisfy
\begin{equation}\label{eq:def_vs}
M_\pm(\ga)v_s^\pm(\ga)=\mp\ga v_s^\pm(\ga).
\end{equation}
Using the definition of $\ga$ from equation (\ref{eq:def_ga}),
the Evans function on the Riemann surface is given by
\begin{equation}\label{eq:def_E}
E(\ga)=m(\ga,x)\,
 (\bfy_s^-\wedge\bfy_f^-\wedge
  \bfy_s^+\wedge\bfy_f^+)(\ga,x),
\end{equation}
where
\[
\bfy_f^\pm(\ga,x)=
  (\bfy_{f,1}^\pm\wedge\cdots\wedge\bfy_{f,n-1}^\pm)(\ga,x)
\]
and
\[
m(\ga,x)=\exp\left(-\int_0^x\mbox{tr}\,M(\ga,s)\,ds\right).
\]

We make a further assumption to 
allow the possibility of bounded and/or exponentially decaying solutions
to equation (\ref{eq:lin_eval}) at $\la=0$; this is not a restriction,
since we allow $k=0$, 
but simply sets up the notation to handle such solutions.

\begin{ass}\label{ass:bounded_sols} The slow solutions satisfy
$\bfy_s^-(0,x)=\bfy_s^+(0,x)$.
Furthermore, there exists a $k$, 
with $0\le k\le n-1$, 
such that $\bfy_{f,i}^-(0,x)=\bfy_{f,i}^+(0,x)$ for $i=0,\dots,k$.
\end{ass} 

\begin{remark} If $\{0\}\notin\sig_c(L)$, then $k$ would be the
geometric multiplicity of the eigenvalue $\la=0$.
\end{remark}

The functions $\bfy_{f,i}^\pm(\ga,x)$ are analytic in $\la$ at
$\la=0$; hence, their derivatives with
respect to $\ga$ are related to derivatives with respect to $\la$
by the chain rule, and when evaluated at $\la=\ga=0$ satisfy 
\begin{equation}\label{eq:la_ga_rel}
m!\,\partial_\ga^{2m}\bfy_{f,i}^\pm(0,x)=
   \frac{(2m)!}{b'(0)^m}\,\partial_\la^m\bfy_{f,i}^\pm(0,x).
\end{equation}
The solutions $\bfy_s^\pm(\ga,x)$ are not analytic in $\la$ at
$\la=0$; however, by the assumptions on the eigenvalues of $M_\pm(\la)$
they are analytic in $\ga$ (\cite{markushevich:tof85}).
Since $\bfy_s^-(0,x)=\bfy_s^+(0,x)$, we have  
$E(0)=0$ from (\ref{eq:def_E}). 
As a consequence of Assumption \ref{ass:bounded_sols} and 
equation (\ref{eq:la_ga_rel}), we expect that
$\partial_\ga^{2k+1}E(0)\neq0$ with $\partial_\ga^jE(0)=0$ for
$0\le j\le 2k$.
Proving this conjecture will be the focus of the next two subsections.

\subsection{Derivatives of the slow components}

The definition of the Evans function in (\ref{eq:def_E})
is based on $2n$ solutions of equation (\ref{eq:lin_eval}). 
We can specify a related set of $2n$ linearly independent solutions 
$\{ \bfu_1,\dots,\bfu_{2n} \}$ to (\ref{eq:lin_eval}) at $\la=0$, 
which are useful for differentiating components of the Evans function, as 
follows. 
Set $\bfu_i(x)=\bfy_{f,i}^-(0,x)$ for $i=1,\dots,k$.
The existence of $k$ independent solutions which grow exponentially fast as
$|x|\to\infty$ is guaranteed by a result in Gardner and Jones
\cite{gardner:two89}; let $\bfu_i(x),\,i=k+1,\dots,2k$ be these
solutions.
Now set
\[
\begin{array}{clll}
\bfu_{2k+i}(x)&=&{\ds\bfy_{f,k+i}^-(0,x)},\quad&i=1,\dots,n-k-1 \\
\vspace{.1mm} \\
\bfu_{n+k-1+i}(x)&=&{\ds\bfy_{f,k+i}^+(0,x)},\quad&i=1,\dots,n-k-1.
\end{array}
\]
Finally, set $\bfu_{2n-1}(x)=\bfy_s^-(0,x)$, and let $\bfu_{2n}(x)$
be chosen so that
\begin{equation}\label{eq:def_u2n}
m(0,x)\,\bfu_1(x)\wedge\cdots\wedge\bfu_{2n}(x)=1.
\end{equation}

Now, $m(0,x)\,
\bfu_1(x)\wedge\cdots\wedge\bfu_{2n-1}(x)$ induces a solution
$\bfu^A_{2n}(x)$ to the adjoint equation associated with equation
(\ref{eq:lin_eval}); furthermore, $\bfu^A_{2n}(x)\cdot\bfu_{2n}(x)=1$
(\cite{alexander:ati90,kapitula:tef99,sandstede:som98}).
In all of the examples having the branch point structure under consideration 
of which the authors are aware, this
particular adjoint solution is bounded above and bounded from zero 
as $|x|\to\infty$; hence, this will be an assumption.
The theory can be appropriately modified if this does not hold true.

\begin{ass}\label{ass:ua2n} There exist positive constants $C_1$ and $C_2$
such that the adjoint solution $\bfu^A_{2n}(x)$ satisfies
$C_1\le |\bfu^A_{2n}(x)|\le C_2$ for all $x\in\R$.
\end{ass}
  
To differentiate the Evans function at $\ga=0$,
it is necessary to derive an expression for 
$\partial_\ga(\bfy_s^- -\bfy_s^+)(0,x)$ at some value of $x$.
Set 
\[
\bfz_s^\pm(\ga,x)=\bfy_s^\pm(\ga,x)e^{\pm\ga x},
\]
and note that for fixed $x$, 
\[
\partial_\ga\bfy_s^\pm(0,x)=\partial_\ga\bfz_s^\pm(0,x).
\]
Following Kapitula and Sandstede \cite{kapitula:sob98}, write
\begin{equation}\label{eq:def_zs}
\bfz_s^\pm(\ga,x)=v_s^\pm(\ga)+\bfy_s^\pm(0,x)-v_s^\pm(0)+w^\pm(\ga,x),
\end{equation}
where $w^\pm(\ga,x)$ is assumed to decay exponentially fast as
$x\to\pm\infty$ and to satisfy $w^\pm(0,x)=0$.
This ansatz is valid due to equation (\ref{eq:def_ys}).

The assumption that $b'(0)\neq0$ implies that we can locally
write $\la=b^{-1}(\ga^2)$, which yields that $d\la/d\ga=0$ at $\ga=0$.
Since $M(\la,x)$ is analytic in $\la$, we then observe that
$\partial_\ga M(0,x)=0$.
Therefore, it can be readily seen that
\begin{equation}\label{eq:wpm}
\begin{array}{lll}
\partial_x(\partial_\ga w^\pm(0,x))&=&M(0,x)\partial_\ga w^\pm(0,x)\\
\vspace{.1mm} \\
&&\quad\quad +M(0,x)\partial_\ga v_s^\pm(0)\pm\bfy_s^\pm(0,x).
\end{array}
\end{equation}
The nonhomogeneous term in the above equation decays exponentially
fast as $x\to\pm\infty$.
This can be seen by noting that as a consequence of equation
(\ref{eq:def_vs}), $M_\pm(0)\partial_\ga v_s^\pm(0)=\mp v_s^\pm(0)$.

Set
\[
\bfg^\pm(x)=M(0,x)\partial_\ga v_s^\pm(0)\pm\bfy_s^\pm(0,x).
\]
Solving equation (\ref{eq:wpm}) with variation of parameters 
(see \cite{kapitula:tef99}) yields 
\begin{equation}\label{eq:wpm_sol_temp}
\begin{array}{lll}
\partial_\ga w^\pm(0,0)&=&{\ds\sum_{i=1}^{n-1}c_i^\pm\bfy_{f,i}^\pm(0,0)
 +c_s^\pm\bfy_s^\pm(0,0)} \\
\vspace{.1mm} \\
&&\quad {\ds +\sum_{i=k+1}^{2k}\bfu_i(0)\int_{\pm\infty}^0
  \bfg^\pm(x)\cdot\bfu^A_i(x)\,dx} \\
\vspace{.1mm} \\
&&\quad\quad {\ds+\bfu_{2n}(0)\int_{\pm\infty}^0\bfg^\pm(x)
     \cdot\bfu^A_{2n}(x)\,dx}. 
\end{array}
\end{equation}
Here $\bfu^A_i(x)$ are solutions to the adjoint equation associated
with equation (\ref{eq:lin_eval}) satisfying $\bfu^A_i(x)\cdot\bfu_j(x)
=\del_{ij}$, and $c^\pm_i$ are some constants. 
As a consequence of the manner in which the solutions $\bfu_i(x)$
were defined, $\bfu^A_i(x)$ decays exponentially fast as $|x|\to\infty$
for $i=k+1,\dots,2k$; hence, the improper integrals are valid.
The observation that 
\[
M(0,x)\partial_\ga v_s^\pm(0)\cdot\bfu^A_i(x)=
  -\partial_\ga v_s^\pm(0)\cdot\frac{d}{dx}\bfu^A_i(x)
\]
together with the exponential decay of the adjoint solutions $\bfu^A_i(x)$ 
simplify the solution formula in equation (\ref{eq:wpm_sol_temp}) to 
\begin{equation}\label{eq:wpm_sol}
\begin{array}{lll}
\partial_\ga w^\pm(0,0)&=&{\ds\sum_{i=1}^{n-1}c_i^\pm\bfy_{f,i}^\pm(0,0)
 +c_s^\pm\bfy_s^\pm(0,0)} \\
\vspace{.1mm} \\
&&\quad {\ds -\sum_{i=k+1}^{2k}\left[\partial_\ga v_s^\pm(0)
        \cdot\bfu^A_i(0)\right]\,\bfu_i(0)} \\
\vspace{.1mm} \\
&&\quad\quad {\ds+\left[\partial_\ga v_s^\pm(0)\cdot
             (\bfu^A_{2n}(\pm\infty)-\bfu^A_{2n}(0))\right]\,\bfu_{2n}(0)}.
\end{array}
\end{equation} 
Here we note that since $\bfy_s^\pm(0,x)=\bfu_{2n-1}(x),\,\bfy_s^\pm(0,x)
\cdot\bfu_j^A(x)=0$ for $j\neq 2n-1$.
Therefore,  upon an appropriate renaming of the constants one sees that
\begin{equation}\label{eq:wpm_diff}
\begin{array}{l}
\partial_\ga (w^- 
-w^+)(0,0)={\ds\sum_{i=1}^{n-1}\tilde{c}_i^\pm\bfy_{f,i}^\pm(0,0)
 +\tilde{c}_s\bfy_s^-(0,0)} \\
\vspace{.1mm} \\
\quad\quad {\ds+\sum_{i=k+1}^{2k}\left[\partial_\ga (v_s^+ -v_s^-)(0)
     \cdot\bfu^A_i(0)\right]\,\bfu_i(0)} \\
\vspace{.1mm} \\
\quad\quad    {\ds +\left[\partial_\ga (v_s^+ -v_s^-)(0)
     \cdot\bfu^A_{2n}(0)\right]\,\bfu_{2n}(0)} \\
\vspace{.1mm} \\
\quad\quad {\ds+\left[
   \partial_\ga v_s^-(0)\cdot\bfu^A_{2n}(-\infty)
   -\partial_\ga v_s^+(0)\cdot\bfu^A_{2n}(+\infty)\right]\,\bfu_{2n}(0)}.
\end{array}
\end{equation}
The following lemma has now almost been proved.

\begin{lemma}\label{lem:slow_gam_deriv} Suppose that Assumptions
\ref{ass:M}, \ref{ass:bounded_sols}, and \ref{ass:ua2n} hold.
The solutions $\bfy_s^\pm(\ga,x)$ then satisfy
\[
\begin{array}{l}
{\ds \partial_\ga(\bfy_s^- -\bfy_s^+)(0,0)=
\sum_{i=1}^{n-1}c_i^\pm\bfy_{f,i}^\pm(0,0)+c_s\bfy_s^-(0,0)} \\
\vspace{.1mm} \\
\quad\quad+{\ds\left[
   \partial_\ga v_s^-(0)\cdot\bfu^A_{2n}(-\infty)
   -\partial_\ga v_s^+(0)\cdot\bfu^A_{2n}(+\infty)\right]\,\bfu_{2n}(0)} \\
\end{array}
\]
for some constants $c_i^\pm,\,c_s$.
\end{lemma}

\proof As a consequence of equation (\ref{eq:def_zs}), it follows  
that
\[
\partial_\ga(\bfy_s^- -\bfy_s^+)(0,0)=\partial_\ga(v_s^- -v_s^+)(0)
  +\partial_\ga (w^- -w^+)(0,0),
\]
where $\partial_\ga (w^- -w^+)(0,0)$ is given in equation
(\ref{eq:wpm_diff}).
Plugging in the fact that 
\[
\partial_\ga(v_s^- -v_s^+)(0)=\sum_{i=1}^{2n}
 \left[\partial_\ga(v_s^- -v_s^+)(0)\cdot\bfu^A_i(0)\right]\,\bfu_i(0)
\]
therefore yields the result.
\qed

\begin{remark}\label{rem:weak_ass} If only one of the matrices 
$M_\pm(\la)$, say $M_-(\la)$,
satisfies Assumption \ref{ass:M}, i.e., the other matrix, say $M_+(\la)$,
is such that all of its eigenvalues are analytic in $\la$ at $\la=0$, then
it is only necessary to compute the relevant term
$\partial_\ga \bfy_s^-(0,0)$.
One can then drop the term $\partial_\ga v_s^+(0)\cdot\bfu^A_{2n}(+\infty)$
in the above lemma.
\end{remark} 

%%%%%%%%%%%%%%%%%%%%%%%%%%%%%%%%%%%%%%%%%%%%%%%%%%%%%%%%%%%%%%%%%%%%%%
\subsection{Derivatives of the Evans function}

We are now ready to derive expressions for certain derivatives of the Evans 
function
with respect to $\ga$ at $\ga=0$. 
Recall Assumption \ref{ass:bounded_sols}, which states that there
exist $k$ solutions at $\la=0$ to equation (\ref{eq:lin_eval}) which
decay exponentially as $|x|\to\infty$.
By the construction of the system (\ref{eq:lin_eval}) it must then be true that 
for $i=1,\dots,k$
\[
\bfy_{f,i}^\pm(0,x)=[\psi_{1,i},\psi_{1,i}']^T,
\]
where $L\psi_{1,i}=0$.
We assume that although $\la=0$ is not an isolated eigenvalue of 
finite multiplicity, we can nonetheless find ``generalized eigenfunctions''
for $\la=0$.  

\begin{ass}\label{ass:J_chain}  There exist numbers $a_i$ and 
functions  
$\psi_{j,i},\,i=1,\dots,k,\,j=1,\dots,a_i,$ such that
\[
L\psi_{j,i}=\psi_{j-1,i},\quad\psi_{0,i}=0.
\]
Furthermore, if $j\ge2$, then $|\psi_{j,i}(x)|$ decays exponentially fast
as $|x|\to\infty$.
\end{ass}

\begin{remark} If $\la=0$ were an isolated eigenvalue with finite
multiplicity, then the exponential decay assumption would hold automatically.
Otherwise, it is possible for the generalized eigenfunctions to either be 
bounded away from zero or even grow like some power of $|x|$ as $|x|\to\infty$
(see Section 3.5).
\end{remark}

Set $p=\sum_{i=1}^k a_i$, and let
\begin{equation}\label{eq:def_bfpsi}
\bfpsi_{a_i,i}(x)=[\psi_{a_i,i},\psi_{a_i,i}']^T
\end{equation}
for $i=1,\dots,k$.
Following Kapitula \cite{kapitula:tef99} it can be shown that 
$\partial_\la^{a}(\bfy_{f,i}^- -\bfy_{f,i}^+)(0,x)=0$ for positive integers 
$a<a_i$, and 
\begin{equation}\label{eq:la_der}
\begin{array}{l}
{\ds \partial_\la^{a_i}(\bfy_{f,i}^- -\bfy_{f,i}^+)(0,x)=
      \sum_{j=1}^{n-1} d_j^\pm\bfy_{f,j}^\pm(0,x) +d_s\bfy_s^-(0,x)}\\
\vspace{.1mm} \\
\quad {\ds +d_{2n}\bfu_{2n}(x)+a_i!\sum_{j=k+1}^{2k} 
     <\partial_\la M(0,x)\bfpsi_{a_i,i}(x),\bfu^A_j(x)>\,
      \bfu_j(x)},
\end{array}
\end{equation}
for constants $d_j^\pm,\,d_s,$ and $d_{2n}$.
In the above,
\[
<\bfg(x),\bfh(x)>=\int_{-\infty}^{+\infty}\bfg(x)\cdot\bfh(x)\,dx.
\]
The integrals are valid due to the fact that the adjoint solutions
decay exponentially fast as $|x|\to\infty$.

Recall the definition of the Evans function given in equation (\ref{eq:def_E}).
As a consequence of the above discussion and equation (\ref{eq:la_ga_rel}), 
$\partial_\ga^{m}E(0)=0$ for any positive integer $m<2p+1$.
Upon using relation (\ref{eq:la_ga_rel}),
differentiation yields 
\[
\begin{array}{lll}
\partial_\ga^{2p+1}E(0)&=&{\ds \frac{(2p+1)!}{\prod_{i=1}^k (2a_i)!}\,
     \partial_\ga(\bfy_s^- -\bfy_s^+)\wedge
 \partial_\ga(\bfy_f^- -\bfy_f^+)\wedge\Phi} \\
\vspace{.1mm} \\
&=&{\ds \frac{(2p+1)!}{b'(0)^p\prod_{i=1}^k a_i!}\,
     \partial_\ga(\bfy_s^- -\bfy_s^+)\wedge
 \partial_\la(\bfy_f^- -\bfy_f^+)\wedge\Phi},
\end{array}
\]
where
\[
\partial_\ga(\bfy_f^- -\bfy_f^+)=
  \partial_\la^{a_1}(\bfy_{f,1}^- -\bfy_{f,1}^+)\wedge\cdots\wedge
 \partial_\la^{a_k}(\bfy_{f,k}^- -\bfy_{f,k}^+),
\]
and
\[
\partial_\la(\bfy_f^- -\bfy_f^+)=
  \partial_\la^{a_1}(\bfy_{f,1}^- -\bfy_{f,1}^+)\wedge\cdots\wedge
   \partial_\la^{a_k}(\bfy_{f,k}^- -\bfy_{f,k}^+),
\]
and
\[
\Phi(x)=m(0,x)\left(\bfy_{f,k+1}^-\wedge\cdots\wedge\bfy_{f,n-1}^-
  \wedge\bfy_s^+\wedge\bfy_{f,k+1}^+\wedge\cdots\wedge
  \bfy_{f,n-1}^+\right).
\]
Substituting the result of Lemma \ref{lem:slow_gam_deriv} and
equation (\ref{eq:la_der}) into this expression, one obtains the
following theorem.

\begin{theorem}\label{thm:ga_der} Suppose that the assumptions leading
to Lemma \ref{lem:slow_gam_deriv} hold, and that Assumption \ref{ass:J_chain} 
holds.
Then derivatives of the Evans function defined from the linear operator
$L$ satisfy 
\[
\partial_\ga^{2p+1}E(0)=-\frac{(2p+1)!}{b'(0)^p}\,\al\,D 
\]
where
\[
\al=\partial_\ga v_s^-(0)\cdot\bfu^A_{2n}(-\infty)
   -\partial_\ga v_s^+(0)\cdot\bfu^A_{2n}(+\infty)
\]
and
\[
D=\left|\begin{array}{ccc}
  <\partial_\la M\bfpsi_{a_1,1},\bfu^A_{k+1}>&\cdots&
        <\partial_\la M\bfpsi_{a_1,1},\bfu^A_{2k}> \\
   \vdots&{}&\vdots \\
  <\partial_\la M\bfpsi_{a_k,k},\bfu^A_{k+1}>&\cdots&
       <\partial_\la M\bfpsi_{a_k,k},\bfu^A_{2k}>
                                          \end{array}\right|.
\]
\end{theorem}

\begin{remark}\label{rem:ga_der} A similiar theorem was proved in Kapitula
\cite{kapitula:tef99} in the case that $\la=0$ is an
isolated eigenvalue with finite multiplicity.
\end{remark}

\begin{remark}
Another case that may arise is that $b(0)=b'(0)=0$.
Since $b(\la)$ is analytic, similar expressions for the derivatives of $E(\ga)$
at $\ga=0$
can be derived via the chain rule; the more zero derivatives 
$b(\la)$ has, the more complicated the results.
Such an example arises in Section 3.5.
\end{remark}

%%%%%%%%%%%%%%%%%%%%%%%%%%%%%%%%%%%%%%%%%%%%%%%%%%%%
\subsection{Example: CGL}

Consider the linearized problem for the CGL (\ref{eq:GL}), 
given in Section 4 in equation (\ref{eq:L_full}).
Upon setting $\ep=0$, the matrix $M_0(\la,x)$ is given by
\begin{equation}\label{eq:def_Mgl}
M_0(\la,x)=\left[\begin{array}{cccc}0&0&1&0 \\ 0&0&0&1 \\
2(\la-1+3\Phi^2) &0&0&0 \\ 0& 2(\la-1+\Phi^2) &0&0 \end{array}\right].
\end{equation}
It is easy to check here that $b(\la)=2\la$.
Following the procedure leading up to equation (\ref{eq:def_u2n}),
choose the solutions to $\bfy'=M_0(0,x)\bfy$ to be
\begin{equation}\label{eq:def_ui}
\begin{array}{llllll}
\bfu_1&=&[\Phi',0,\Phi'',0]^T,\,\,
  &\bfu_2&=&[u_2^1,0,u_2^3,0]^T \\
\vspace{.1mm} \\
\bfu_3&=&[0,\Phi,0,\Phi']^T,\,\,
  &\bfu_4&=&[0,u_4^2,0,u_4^4]^T
\end{array}
%\bfu_1=\left[\begin{array}{c}\Phi'\\0\\
%     \Phi''\\0\end{array}\right],\,\,
%\bfu_2=\left[\begin{array}{c}u_2^1\\0\\
%     u_2^3\\0\end{array}\right],\,\,
%\bfu_3=\left[\begin{array}{c}0\\ \Phi\\0
%     \\ \Phi'\end{array}\right],\,\,
%\bfu_4=\left[\begin{array}{c}0\\ u_4^2\\0
%     \\ u_4^4\end{array}\right]
\end{equation}
($u_4^2(x)=x\Phi(x)-1,\,u_4^4(x)=\Phi(x)+x\Phi'(x)$).
The solution $\bfu_2$, which grows exponentially fast as $x\to\pm\infty$, 
is chosen so that
\[
\left|\begin{array}{cc}\Phi' & u_2^1 \\ \Phi'' & u_2^3\end{array}
   \right|=-1;
\]
hence, $\bfu_1,\cdots,\bfu_4$ satisfies (\ref{eq:def_u2n}).
While it is possible to find an explicit expression for $\bfu_2$, it
is not necessary, and hence will not be done.
The adjoint solutions satisfying $\bfu_i\cdot\bfu^A_j=\del_{ij}$ are
then given by
\begin{equation}\label{eq:def_uai}
\begin{array}{llllll}
\bfu^A_1&=&[-u_2^3,0,u_2^1,0]^T,\,\,
   &\bfu^A_2&=&[\Phi'',0,-\Phi',0]^T \\
\vspace{.1mm} \\
\bfu^A_3&=&[0,u_4^4,0,-u_4^2]^T,\,\,
   &\bfu^A_4&=&[0,-\Phi',0,\Phi]^T.
\end{array}
%\bfu^A_1=\left[\begin{array}{c}-u_2^3\\0\\
%     u_2^1\\0\end{array}\right],\,\,
%\bfu^A_2=\left[\begin{array}{c}\Phi''\\0\\
%     -\Phi'\\0\end{array}\right],\,\,
%\bfu^A_3=\left[\begin{array}{c}0\\ u_4^4\\0
%     \\ -u_4^2\end{array}\right],\,\,
%\bfu^A_4=\left[\begin{array}{c}0\\ -\Phi'\\0
%     \\ \Phi\end{array}\right].
\end{equation}

Under the normalization $\bfy_s^\pm(0,x)=\bfu_3(x)$, a simple calculation 
reveals
that
\begin{equation}\label{eq:def_vsgl}
v_s^\pm(\ga)=[0,\pm 1,0,-\ga]^T
\end{equation}
(recall that $\ga^2=2\la$ in this case).
The result of Theorem \ref{thm:ga_der}, with $a_1=1$ and
$\bfpsi_{1,1}=\bfu_1$, then implies that
\[
\alpha = \partial_\ga v_s^-(0)\cdot\bfu^A_{4}(-\infty)
   -\partial_\ga v_s^+(0)\cdot\bfu^A_{4}(+\infty)=2,
\]
and hence
\begin{equation}\label{eq:E_ggg}
\begin{array}{lll}
\partial_\ga^3 E(0)&=&{\ds12\int_{-\infty}^{+\infty}(\Phi')^2(x)\,dx} \\
\vspace{.1mm} \\
&=&16.
\end{array}
\end{equation}

The linearized eigenvalue problem when $\ep=0$ can be
written as
\[
L_+p=\la p,\quad L_-q=\la q,
\]
where $L_\pm$ are defined in equation (\ref{eq:def_Lpm}).
As such, we can actually say much more about the Evans function.
First, both operators $L_\pm$ are self-adjoint, so their
spectra must be real.
Furthermore, since $L_+\Phi'=0$ and $\Phi'$ has no zeros, an application
of St\"urm-Louiville theory implies that $\la=0$ is
the largest eigenvalue for $L_+$.
Similiarly, there are no positive eigenvalues for $L_-$.
Therefore, the following lemma holds for the Evans function.

\begin{lemma}\label{lem:evans_gl} Suppose that $\ep=0$.
Set $\ga^2=2\la$.
For $\ga$ near zero the Evans function has the expansion
\[
E(\ga)=\frac83\ga^3+O(\ga^4).
\]
Furthermore, the Evans function is nonzero for $\Re\ga>0$.
\end{lemma}

\begin{remark} As a consequence of this lemma, for a perturbed
problem it suffices to locate the zeros of the Evans function near
$\ga=0$ to determine the stability of the wave.
\end{remark}

%%%%%%%%%%%%%%%%%%%%%%%%%%%%%%%%%%%%%%%%%%%%%%%%%%%%%%%%%%%%%%%%%%%%
\subsection{Example: NLS}

Consider the linearized problem for the PNLS (\ref{eq:NLS_pert}),
given in Section 5 in equation (\ref{eq:L_full_NLS}).
Upon setting $\ep=0$, the matrix $M_0(\la,x)$ is given by
\begin{equation}\label{eq:def_Mgl_nls}
M_0(\la,x)=\left[\begin{array}{cccc}0&0&1&0 \\ 0&0&0&1 \\
2(-1+3\Phi^2) &-2\la&0&0 \\ 2\la& 2(-1+\Phi^2) &0&0 \end{array}\right].
\end{equation}
Choose the solutions $\bfy'=M_0(0,x)\bfy$ to be those given in
equation (\ref{eq:def_ui}), and let the adjoint solutions be those given
in equation (\ref{eq:def_uai}).
Define $\ga$ by
\begin{equation}\label{eq:def_ga_nls}
\ga^2=2(1-\sqrt{1-\la^2}),
\end{equation}
so that upon taking the principal square root,
\[
\la=\frac12\ga\sqrt{4-\ga^2}.
\]
Note that
\[
\la=\ga+O(\ga^2)
\]
for $\ga$ sufficiently small, so that
\[
\frac{\partial}{\partial\la}=\frac{\partial}{\partial\ga}
\]
at $(\la,\ga)=(0,0)$.
Under the normalization $\bfy_s^\pm(0,x)=\bfu_3(x)$, a simple calculation 
reveals
that
\begin{equation}\label{eq:def_vsnls}
v_s^\pm(\ga)=-\frac12\,[\mp\ga,\mp\sqrt{4-\ga^2},\ga^2,\ga\sqrt{4-\ga^2}]^T.
\end{equation}
Thus, the result of Lemma \ref{lem:slow_gam_deriv} implies 
that
\begin{equation}\label{eq:bfys_NLS}
\partial_\ga(\bfy_s^- -\bfy_s^+)(0,0)=2\bfu_4(0)+c_1\bfu_1(0)+c_3\bfu_3(0).
\end{equation}

In this example, $b(\la)$ is given in (\ref{eq:def_ga_nls}), so $b(0)=0$,
 but $b'(0)=0$ as well.
As noted in Remark \ref{rem:ga_der}, this does not in itself rule out use of a 
modified
form of Theorem \ref{thm:ga_der}. 
Unfortunately, the result of Theorem \ref{thm:ga_der} truly cannot be applied 
here.
Since the generalized eigenfunctions are given by
\[
\psi_{1,2}(x)=\left[\begin{array}{c}0\\ \Phi(x)\end{array}\right],\quad
\psi_{2,2}(x)=\frac12\left[\begin{array}{c}x\Phi'(x)+\Phi(x) 
\\0\end{array}\right],
\]
the assumption that the generalized eigenfunctions decay exponentially
fast as $|x|\to\infty$ does not hold.
Thus, we must construct the desired solutions directly.
Using the fact that
\[
(\partial_\la\bfy_f^\pm)'=M_0\partial_\la\bfy_f^\pm
    +\partial_\la M_0\bfy_f^\pm,
\]
and that $\bfy_f^\pm(0,x)=\bfu_1(x)$, it is not hard to verify that
\begin{equation}\label{eq:bfyf_NLS}
\partial_\la\bfy_f^\pm(0,x)=-\bfu_4(0)\mp\bfu_3(x).
\end{equation}
Thus, upon solving the equation
\[
(\partial_\la^2\bfy_f^\pm)'=M_0\partial_\la^2\bfy_f^\pm
    +2\partial_\la M_0\partial_\la\bfy_f^\pm
\]
by variation of parameters, one finds that
\[
\partial_\la^2(\bfy_f^- -\bfy_f^+)(0,0)=4\bfu_2(0)+c_1\bfu_1(0).
\]
Combining this result with equation (\ref{eq:bfys_NLS}) implies that 
when $\ep=0$,
\begin{equation}\label{eq:E_ggg_NLS}
\begin{array}{lll}
\partial_\ga^3 E(0)&=&3\,\partial_\ga(\bfy_s^- -\bfy_s^+)\wedge
    \partial_\ga^2(\bfy_f^- -\bfy_f^+)\wedge\bfy_s^+\wedge\bfy_f^+\\
\vspace{.1mm} \\
&=&-24.
\end{array}
\end{equation}
The following lemma is now almost proved.

\begin{lemma}\label{lem:evans_nls} Suppose that $\ep=0$.
Set $\ga^2=2(1-\sqrt{1-\la^2})$.
For $\ga$ near zero the Evans function has the expansion
\[
E(\ga)=-4\ga^3+O(\ga^4).
\]
Furthermore, the Evans function is nonzero for $\Re\ga\ge0$
except at $\ga=0$.
\end{lemma}

\proof It is shown in Chen \etal \cite{chen:adp98} that the squared
Jost solutions of the Zakharov-Shabat eigen-equation, i.e., the
squared eigenfunctions, form a complete set.
In other words, bounded eigenfunctions for the
linearized problem exist if and only if $\la\in i\R$ (or $\ga\in i\R$).
Thus, the Evans function is nonzero for $\Re\ga>0$, and to complete
the proof we must show that it is nonzero on the set
$i\R\backslash\{0\}$.

To this end, we will rewrite the eigenvalue problem in such a way
as to fully exploit the results presented in \cite{chen:adp98}.
Letting $\psi=\phi^*$, the NLS can be rewritten as the system
\[
\begin{array}{rcl}
i\phi_t-\frac12\phi_{xx}-\phi+\phi^2\psi&=&0 \\
\vspace{.1mm} \\
-i\psi_t-\frac12\psi_{xx}-\psi+\phi\psi^2&=&0.
\end{array}
\]
Linearizing about the wave $\Phi$ yields the system
\[
\begin{array}{rcl}
i\phi_t-\frac12\phi_{xx}-\phi+2\Phi^2\phi+\Phi^2\psi&=&0 \\
\vspace{.1mm} \\
-i\psi_t-\frac12\psi_{xx}-\psi+\Phi^2\phi+2\Phi^2\psi&=&0,
\end{array}
\]
which, upon setting
\[
(\phi,\psi)\to(\phi,\psi)e^{i\rho t},
\]
induces the eigenvalue problem
\[
\begin{array}{llr}
\frac12\phi''+(1-2\Phi^2)\phi-\Phi^2\psi&=&-\rho\phi \\
\vspace{.1mm} \\
\frac12\psi''+(1-2\Phi^2)\psi-\Phi^2\phi&=&\rho\psi
\end{array}
\]
($'=d/dx$).

Since $\ga\in i\R$ if and only if $\rho\in\R$, we will now explicitly
construct the Evans function for real $\rho$.
In the usual way, the eigenvalue system
\[
\bfy'=M(\rho,x)\bfy
\]
can be constructed.
Set
\[
\xi=\rho+\sqrt{1+\rho^2},
\]
where the principal square root is taken.
Note that $\rho\in\R$ implies that $\xi\in\R^+$, and that $\rho=0$ implies
that $\xi=1$.
The eigenvalues for the asymptotic matrix $M_0(\xi)$ are given by
$\pm\mu_f(\xi),\,\pm\mu_s(\xi)$, where
\[
\mu_f(\xi)=\frac{\xi+1}{\sqrt{\xi}},\quad
\mu_s(\xi)=i\frac{\xi-1}{\sqrt{\xi}},
\]
and the principal square root is being taken.
The corresponding eigenvectors are given by
\[
v_f^\pm=[1,\xi,\pm\mu_f,\pm\xi\mu_f]^T,\quad
v_s^\pm=[1,-1/\xi,\pm\mu_s,\mp\mu_s/\xi]^T.
\]

Now, when $\Re\ga>0,\,\Im\rho<0$, so that for $\Im\xi\le0$ we need
to define the solutions $\bfy_s^\pm$ and $\bfy_f^\pm$ comprising the 
Evans function so that
\[
\lim_{x\to\pm\infty}(\bfy_s^\pm\wedge\bfy_f^\pm)(\xi,x)e^{\pm(\mu_s+\mu_f)x}=
   v_s^\mp\wedge v_f^\mp.
\]
This is done so that the definition of the Evans function is consistent
with that given in equation (\ref{eq:def_E}).
Using the information presented in \cite{chen:adp98}, it can readily
be checked that
\[
\lim_{x\to+\infty}(\bfy_s^-\wedge\bfy_f^-)(\xi,x)e^{-(\mu_s+\mu_f)x}=
   a(\xi)b(\xi)\,v_s^+\wedge v_f^+,
\]
where
\[
a(\xi)=\frac{\sqrt{\xi}-i}{\sqrt{\xi}+i},\quad
b(\xi)=\left(\frac{\sqrt{\xi}-1}{\sqrt{\xi}+1}\right)^2.
\]
Thus, we get that
\[
\begin{array}{lll}
E(\xi)&=&{\ds \lim_{x\to+\infty}(\bfy_s^-\wedge\bfy_f^-\wedge\bfy_s^+
    \wedge\bfy_f^+)(\xi,x) }  \\
\vspace{.1mm} \\
&=&a(\xi)b(\xi)\,v_s^-\wedge v_f^-\wedge v_s^+\wedge v_f^+.
\end{array}
\]
Since
\[
v_s^-\wedge v_f^-\wedge v_s^+\wedge v_f^+=-4i\frac{(1+\xi^2)^2(1-\xi^2)}
  {\xi^3},
\]
we see that $E(\xi)\neq0$ for $\xi\in\R^+$ except when $\xi=1$.
As $\xi=1$ corresponds to $\rho=0$, the proof is complete.
\qed

\begin{remark} The functions $a(\xi)$ and $b(\xi)$ are related to
the transmission coefficient for the Zakharov-Shabat inverse scattering
problem.
\end{remark}

\begin{remark} As a consequence of Proposition 2.17 in \cite{kapitula:sob98},
the Evans function will remain nonzero for $\ep>0$ and $|\ga|$ sufficiently
large.
Therefore, for a perturbed
problem it suffices to locate the zeros of the Evans function near
$\ga=0$ to determine the stability of the wave.
\end{remark}

\section{Perturbation calculations at the branch point: CGL}
\setcounter{equation}{0}

In the next two sections we will be using the Evans function to
locate the eigenvalues that bifurcate out of the branch point.
To accomplish this task, we will need to perform
perturbation calculations for the various coefficients
of terms in the series expansions for the Evans function.
Fortunately, the techniques have been developed that will enable
us to do so.
In Kapitula \cite{kapitula:tef99}, a procedure was described which
allows one to perform these calculations for expansions about an
eigenvalue that is isolated with finite multiplicity.
This assumption is not valid for the systems
considered in this paper, as we wish to do
perturbation calculations around a branch point; however, all is
not lost.
Kapitula and Sandstede \cite{kapitula:sob98} showed that it is
possible to do perturbation calculations around a branch point
if a transformation is done on the eigenvalue parameter so
that the branch point does not move under the perturbation.
By combining and appropriately modifying the approaches of these two 
works, together with the results in Section 3, we are able to do an expansion 
around the branch point in 
terms of the transformed eigenvalue parameter. 
Recall the manner in which $E(\ga)$ is defined in equation (\ref{eq:def_E}).
To compute the coefficients in the Taylor expansion for $E(\ga)$, we will
need to be able to compute terms such as $\partial_\ep^k
(\bfy_f^- -\bfy_f^+)(0,0)$ for an appropriate value of $k$.
The first three subsections are devoted to this task.

Henceforth, set 
\begin{equation}{\label{eq:def_ag}}
\Gamma=d_1+d_3+2d_4, \; a=\Gamma\psi_{\ep}^+, 
\end{equation}
where $\psi_{\ep}^+$ is specified by (\ref{eq:def_psi+}) and (\ref{eq:psi_ep}). 
Note that $a$ is exactly the parameter that appears on the left hand
side of conditions (\ref{eq:regpert}) and (\ref{eq:regpert2}); 
that is, the sign of $a$ is directly related to the structure of the 
manifolds whose intersection forms the hole solution. 
 
\subsection{Preliminaries}

After setting $\phi=u+iv$ in equation (\ref{eq:GL}), let the 
perturbation of the wave be written in the form
\[
u+iv=(r+(p+iq))e^{i\int_0^x\psi(s)\,ds}
\]
(this follows the scheme used in Kapitula \cite{kapitula:sow91}).
Here $r$ and $\psi$ are given in Lemma \ref{lem:reg_pert}.
For $\ep\neq0$, the linearized eigenvalue problem derived 
from equation (\ref{eq:GL}), is given, up to $O(\ep^2)$, by 
\begin{equation}\label{eq:L_full}
\la\left[\begin{array}{cc}1-\ep^2d_1^2 & \ep d_1 \\
     -\ep d_1 & 1-\ep^2d_1^2 \end{array}\right]=L_0+\ep L_\ep
    +\frac12\ep^2 L_{\ep\ep},
\end{equation}
where
\begin{equation}\label{eq:def_L0}
L_0=\left[\begin{array}{cc}L_+ & 0 \\ 0 & L_- \end{array}\right]
\end{equation}
with
\begin{equation}\label{eq:def_Lpm}
L_+=\frac12\partial_x^2+1-3\Phi^2,\quad L_-=\frac12\partial_x^2+1-\Phi^2,
\end{equation}
and
\begin{equation}\label{eq:def_Lep}
\begin{array}{lll}
{\ds L_\ep}&=&{\ds -(\psi_\ep\partial_x-\frac{\Phi'}{\Phi}\psi_\ep)
   \left[\begin{array}{rr}0&1\\-1&0\end{array}\right]} \\
\vspace{.1mm} \\
&&\quad\quad {\ds  -2\Phi^2(d_1+d_3+2d_4\Phi^2)
     \left[\begin{array}{rr}0&0\\-1&0\end{array}\right]},
\end{array}
\end{equation}
and 
\begin{equation}\label{eq:def_Lep2}
\begin{array}{lll}
L_{\ep\ep}&=&{\ds -\left[\begin{array}{cc} 6\Phi r_{\ep\ep}+\psi_\ep^2&0 \\
     0 & 2\Phi r_{\ep\ep}+\psi_\ep^2 \end{array}\right]}  \\
\vspace{.1mm} \\
&&\quad\quad {\ds +2d_1(\psi_\ep\partial_x-\frac{\Phi'}{\Phi}\psi_\ep) 
    \left[\begin{array}{cc}1&0\\0&1\end{array}\right]} \\
\vspace{.1mm} \\
&&\quad\quad {\ds+4d_1\Phi^2(d_1+d_3+2d_4\Phi^2)
      \left[\begin{array}{cc}1&0\\0&0\end{array}\right]}.
\end{array}
\end{equation}
Note that
\[
L_+\Phi'=0,\quad L_-\Phi=0.
\]
In the above, $\Phi$ is again given by equation (\ref{eq:def_Phi}).

In the standard way, the expansion for the linear operator $L$ given in
equations (\ref{eq:L_full})-(\ref{eq:def_Lep2}) yields an expansion for 
the matrix $M(\la,x)$, i.e., $M=M_0+M_\ep\ep+M_{\ep\ep}\ep^2/2$.
It is clear that $M(\la,x)\to M_\pm(\la)$ as $x\to\pm\infty$.
The branch point for the Evans function, $\la_b$, is the $\la$ value 
such that the matrices $M_\pm(\la_b)$ have an eigenvalue $\al_\pm^b$ which
has geometric multiplicity one and algebraic multiplicity two.
A routine calculation yields the following proposition.

\begin{prop}\label{prop:branch_pt} For $a$ given by (\ref{eq:def_ag}),  
the branch point of the Evans function is given by
\[
\la_b=-\frac12a^2\ep^4.
\]
Set
\[
\ga=\sqrt{2(\la-\la_b)}.
\]
For $\la$ close to $\la_b$ the eigenvalues of $M_\pm(\la)$ that have 
geometric multiplicity one and algebraic multiplicity two when 
$\la=\la_b$ are given by
\[
\mp\ga+\al_\pm^b,
\]
where
\[
\al_\pm^b=\pm a\ep^2.
\]
When $\la=\la_b$, the associated eigenvectors are given by
\[
\eta_\pm^b=\mp\bfu_4(0)+a\ep^2\bfu_3(0).
\]
\end{prop}

\begin{remark} It should be noted that the location of the branch point
does not depend on which of $M_\pm(\la)$ is being discussed.
\end{remark}

\subsection{Calculations for $\bfy_f^\pm$}

Since $\bfy_f^\pm(\la,x)$ are analytic in an $O(1)$ neighborhood of
the origin, for fixed $x$ these functions have Taylor expansions. 
Together with Proposition \ref{prop:branch_pt}, this implies that 
\begin{equation}\label{eq:fast_full_taylor}
(\bfy_f^- -\bfy_f^+)(\la_b,0)=(\bfy_f^- -\bfy_f^+)(0,0)
 +\partial_\la(\bfy_f^- -\bfy_f^+)(0,0)\la_b+O(\ep^8).
\end{equation}
The behavior of these solutions at  $\la=0$ is fairly well understood.
As a consequence of the derivative formula (\ref{eq:la_der}),
\begin{equation}\label{eq:fast_1der_taylor}
\begin{array}{lll}
\partial_\la(\bfy_f^- -\bfy_f^+)(0,0)&=&
    {\ds<\partial_\la M(0,x)\bfu_1(x),\bfu^A_2(x)>\,
      \bfu_2(0)} \\
\vspace{.1mm} \\
&&\quad\quad {\ds + c\bfu_1(0) +O(\ep)}  \\
\vspace{.1mm} \\
  &=&{\ds -\frac83\,\bfu_2(0) + c\bfu_1(0) 
      +O(\ep)}.
\end{array}
\end{equation}
for some constant $c$.
In addition, since
\begin{equation}{\label{eq:yfpm}} 
\bfy_f^\pm(0,x)=\left[\begin{array}{c}r'(x)\\ (r\psi)(x) 
    \\r''(x)\\ (r\psi)'(x) \end{array}\right]
    \mp\psi_+\left[\begin{array}{c}0\\r(x)\\0\\r'(x)\end{array}\right],
\end{equation} 
where 
\[
\psi_+=\lim_{x\to+\infty}\psi(x),
\]
it is seen that
\begin{equation}\label{eq:fast_0der_taylor}
(\bfy_f^- -\bfy_f^+)(0,0)=2\psi_+
   \left[\begin{array}{c}0\\r(0)\\0\\r'(0)\end{array}\right].
\end{equation}
Since $r(0)=0$ for all $\ep\ge0$, it is necessarily true that
$(\bfy_f^- -\bfy_f^+)(0,0)$ will be a multiple of $\bfu_3(0)$ for
all $\ep\ge0$, and hence it will not make a contribution in the resulting
perturbation calculations for the Evans function.
Since $|\la_b|=O(\ep^4)$, the following lemma has now been proved.

\begin{lemma}\label{lem:yf} The difference in the fast solutions 
satisfies, to leading order, 
\[
\partial_\ep^4(\bfy_f^- -\bfy_f^+)(\la_b,0)=32\,a^2\bfu_2(0)
  +c_{14}\bfu_1(0)+c_{34}\bfu_3(0),
\]
for some constants $c_{14}$ and $c_{34}$.
Furthermore, 
\[
\partial_\ep^j(\bfy_f^- -\bfy_f^+)(\la_b,0)=c_{1j}\bfu_1(0)+c_{3j}\bfu_3(0),
  \quad j=0,\dots,3
\]
for some constants $c_{1j}$ and $c_{3j}$.
\end{lemma}

\subsection{Calculations for $\bfy_s^\pm$}

In this subsection all of the calculations will be performed at
$\ga=0$, where
\begin{equation}\label{eq:def_ga1}
\ga^2=2(\la-\la_b).
\end{equation}
As such, the $\ga$ dependence of solutions will be suppressed.
Set
\[
\bfz_s^\pm(x,\ep)=\bfy_s^\pm(x,\ep)e^{-\al_\pm^b x}.
\]
The rescaled variable then satisfies the ODE
\begin{equation}\label{eq:rescale_ODE}
\partial_x\bfz_s^\pm(x,\ep)=(M(x)-\al_\pm^b\id)\bfz_s^\pm(x,\ep),
\end{equation}
and the asymptotic matrices are now such that they have the Jordan
block $\left[\begin{array}{cc}0&1\\0&0\end{array}\right]$ at $\ga=0$
for all $\ep\ge0$.
Again following the procedure outlined in Kapitula 
and Sandstede \cite{kapitula:sob98}, set
\begin{equation}\label{eq:def_zs1}
\bfz_s^\pm(x,\ep)=\eta_\pm^b(\ep)+\bfy_s^\pm(x,0)
     -\eta_\pm^b(0)+w^\pm(x,\ep),
\end{equation}
where $w^\pm(x,\ep)$ is assumed to decay exponentially fast as 
$x\to\pm\infty$ and satisfy $w^\pm(x,0)=0$.
Furthermore, $w^\pm(x,\ep)$ should not be a scalar multiple of
$\bfu_1(x)$.
The vectors $\eta_\pm^b(\ep)$ are given in Proposition \ref{prop:branch_pt}.
Since $\partial_\ep\eta_\pm^b(0)=\partial_\ep\al_\pm^b=0$, upon
recalling that $M=M_0+M_{\ep}\ep+M_{\ep\ep}\ep^2/2$, it follows that
\begin{equation}\label{eq:wpm1}
\partial_x(\partial_\ep w^\pm(x,0))=M_0(x)\partial_\ep w^\pm(x,0)
 +M_\ep(x)\bfy_s^\pm(x,0),
\end{equation}
and
\begin{equation}\label{eq:wpm2}
\begin{array}{lll}
\partial_x(\partial_\ep^2 w^\pm(x,0))&=&M_0(x)\partial_\ep^2 w^\pm(x,0)
 +M_0(x)\partial_\ep^2\eta_\pm^b \\
\vspace{.1mm} \\
&&\quad +2M_\ep(x)\partial_\ep w^\pm(x,0) \\
\vspace{.1mm} \\
&&\quad
     +(M_{\ep\ep}(x)-\partial_\ep^2\al_\pm^b\id)\bfy_s^\pm(x,0).
\end{array}
\end{equation}

\begin{prop}\label{prop:wpm1} Given the ansatz in equation
(\ref{eq:def_zs1}), the relevant solution to (\ref{eq:wpm1}) satisfies
\[
\partial_\ep w^\pm(x,0)=0.
\]
\end{prop}

\proof This follows immediately from the fact that 
$M_\ep(x)\bfy_s^\pm(x,0)=0$.
\qed

\vspace{3mm}
Upon solving equation (\ref{eq:wpm2}) with the variation of parameters
formulation, and using the facts that 
\[
M_0(x)\partial_\ep^2\eta_\pm^b\cdot\bfu^A_i=
    -\partial_\ep^2\eta_\pm^b\cdot\partial_x\bfu^A_i,
\]
and
\[
M_{\ep\ep}(x)\bfy_s^\pm(x,0)=\Phi(2\Phi r_{\ep\ep}+
   \psi_\ep^2)\,\bfu_3(0),
\]
one obtains  
\[
\begin{array}{l}
\partial_\ep^2(w^- -w^+)(0,0)=[\partial_\ep^2\eta_-^b\cdot\bfu_4^A(-\infty)
  -\partial_\ep^2\eta_+^b\cdot\bfu_4^A(+\infty)]\,\bfu_4(0) \\
\vspace{.1mm}
\quad\quad{\ds
  +\int_{-\infty}^{+\infty}\Phi^2(x)
     (2\Phi(x)r_{\ep\ep}(x)+\psi_\ep^2(x))\,dx\,\bfu_4(0)
  +c\bfu_1(0)}
\end{array}
\]
for some constant $c$.
A tedious calculation reveals that
\[
\int_{-\infty}^{+\infty}\Phi^2(x)
     (2\Phi(x)r_{\ep\ep}(x)+\psi_\ep^2(x))\,dx = -2d_1\psi_\ep^+;
\]
combined with Proposition \ref{prop:branch_pt}, this yields 
the following lemma.

\begin{lemma}\label{lem:ys} The difference in the slow solutions satisfies
\[
\partial_\ep(\bfy_s^- -\bfy_s^+)(0,0)=0,
\]
and
\[
\partial_\ep^2(\bfy_s^- -\bfy_s^+)(0,0)=-4(\frac12d_1+\Ga)\psi_\ep^+\,\bfu_4(0)+
  c_2\bfu_1(0)
\]
for some constants $c_1$ and $c_2$.
\end{lemma}

\proof Following the discussion leading up to the lemma, it is seen that
\[
\partial_\ep^2(w^- -w^+)(0,0)=-4(\frac12d_1+\Ga)\psi_\ep^+\,\bfu_4(0)+
  c\bfu_1(0).
\]
The conclusion now follows from the ansatz given in equation 
(\ref{eq:def_zs1}) and the results of Propositions \ref{prop:branch_pt} and
\ref{prop:wpm1}.
\qed

\subsection{Calculations for the Evans function}

Set
\[
\tilde{\Ga}=(\frac12d_1+\Ga), \quad \tilde{a}=\tilde{\Ga}\psi_\ep^+,
\]
where $\Ga$ is specified by (\ref{eq:def_ag}).
In the sequel, all of the evaluations will be performed at
$(\ga,x,\ep)=(0,0,0)$, and the constants $c_i$ will be unknown
(but irrelevant).

Since $\partial_\ga^2=\partial_\la$, as a consequence of equation
(\ref{eq:fast_1der_taylor}),  
\[
\partial_\ga^2(\bfy_f^- -\bfy_f^+)=
     -\frac83\,\bfu_2 + c_1\bfu_1,
\]
with
\[
\partial_\ga(\bfy_f^- -\bfy_f^+)=0.
\]
Furthermore, as a consequence of Lemma \ref{lem:slow_gam_deriv},
\[
\partial_\ga(\bfy_s^- -\bfy_s^+)=2\bfu_4+c_2\bfu_1+c_3\bfu_3.
\]
From Lemmas \ref{lem:yf} and \ref{lem:ys} one has, respectively, that
\[
\partial_\ep^4(\bfy_f^- -\bfy_f^+)=32a^2\bfu_2
  +c_4\bfu_1+c_5\bfu_3,
\]
and
\[
\partial_\ep^2(\bfy_s^- -\bfy_s^+)=-4\tilde{a}\bfu_4+
  c_6\bfu_1.
\]

We are now in position to write down a perturbation expansion for the
Evans function.
In the following, the $\ep$-dependence of the Evans function is being
implicitly assumed.
First,
\[
\begin{array}{lll}
\partial_\ep^6E(0)&=&{\ds \frac{6!}{2!4!}\partial_\ep^2(\bfy_s^- -\bfy_s^+)
  \wedge\partial_\ep^4(\bfy_f^- -\bfy_f^+)\wedge\bfy_s^+
    \wedge\bfy_f^+ } \\
\vspace{.1mm} \\
&=&{\ds \frac83 6!\,a^2\tilde{a},}
\end{array}
\]
and
\[
\begin{array}{lll}
\partial_\ep^4\partial_\ga E(0)&=&\partial_\ga(\bfy_s^- -\bfy_s^+)
  \wedge\partial_\ep^4(\bfy_f^- -\bfy_f^+)\wedge\bfy_s^+
   \wedge\bfy_f^+  \\
\vspace{.1mm} \\
&=&{\ds -\frac83 4!\,a^2},
\end{array}
\]
and
\[
\begin{array}{lll}
\partial_\ep^2\partial_\ga^2 E(0)&=&\partial_\ep^2(\bfy_s^- -\bfy_s^+)
  \wedge\partial_\ga^2(\bfy_f^- -\bfy_f^+)\wedge\bfy_s^+
   \wedge\bfy_f^+ \\
\vspace{.1mm} \\
&=&{\ds \frac{32}{3}\tilde{a}. }
\end{array}
\]
In addition, recall equation (\ref{eq:E_ggg}), which states that
\[
\partial_\ga^3 E(0)=16.
\]
Note that all lower derivatives of $E$ are zero.
Based on the above expansions, the Evans function can
be written as
\begin{equation}\label{eq:evans_expand_GL}
E(\ga,\ep)=\frac83(\ga^3+\tilde{a}\ep^2\ga^2-a^2\ep^4\ga
  +a^2\tilde{a}\ep^6).
\end{equation}

% Note that $E(-a\ep^2,\ep)>0$ if $a<0$, so that the expansion is
% consistent with the results of Lemmas \ref{lem:evans_la=0} and
% \ref{lem:evans_large_reg}, i.e., the 
% Evans function is positive at $\la=0$ when it is positive for
% large positive real $\la$.

While the zeros of the Evans function can be found analytically, it
is difficult to analyze the resulting expressions.
When $d_1=0$, so that $a=\tilde{a}$, however,  
the roots are given by
\begin{equation}\label{eq:evans_roots}
\ga_1=-1.839\,a\ep^2,\,\,
\ga_{2,3}=(0.420\pm 0.606i)\,a\ep^2.
\end{equation}
Recall that $\ga^2=2(\la-\la_b)$, where $\la_b$ is given in Proposition
\ref{prop:branch_pt}.
The roots of $E(\ga,\ep)$ are valid as eigenvalues if and  
only if $\Re\ga>0$.
This is due to the fact that the sheet $K_0$ of $\Rk$ corresponds to
the principal part of
$\sqrt{2(\la-\la_b)}$.
Thus, if $a>0$, then $\ga_{2,3}$ represent the valid zeros of
the Evans function, while if $a<0$, then $\ga_1$ is the valid zero.
Upon using the inversion formula $\la=\ga^2/2+\la_b$, one has the following
lemma.

\begin{lemma}\label{lem:evans_roots_d1=0} Suppose that $d_1=0$. 
If $a>0$, then the zeros of the Evans function inside the curve $K$ 
are given by
\[
\la_{2,3}=(-0.595\pm0.255i)\,a^2\ep^4.
\]
If $a<0$, then the zero of the Evans function inside $K$ is given by
\[
\la_1=1.191\,a^2\ep^4.
\]
\end{lemma}

\begin{remark} As a consequence, the linearized operator has an unstable
eigenvalue if $a<0$.
\end{remark}

Now suppose that $d_1\neq0$, and set $P_{j1}=d_j/d_1$.
To find the zeros, it is most illustrative to do a standard
bifurcation analysis.
From the definition of $\tilde{a}$, it follows that there is at least
one positive real zero if $(3/2+P_{31}+2P_{41})(1+P_{31}+8P_{41}/5)<0$;
otherwise, there is at least one negative real zero.
In addition, a saddle-node bifurcation occurs on 
the lines
\[
P_{31}+2P_{41}=\mu_{sn}^\pm,
\]
where 
\begin{equation}\label{eq:saddle-node}
\mu_{sn}^\pm=\frac32\frac{\pm\al-2/3}{1\mp\al},\quad
  \al^2=\frac{\sqrt{125}+11}2
\end{equation}
($\mu_{sn}^+=-1.716,\,\mu_{sn}^-=-1.385$).
By checking the sign of $\ga$ when $\partial_\ga E(\ga,\ep)=0$, it is
seen that the zeros created by the saddle-node bifurcation have the
opposite sign from those described above.

If $\psi_\ep^+=0$, then $a=\tilde{a}=0$, so that the branch point does not 
move and the zeros of the Evans function remain at $\ga=0$.
For the rest of the discussion, assume that $\psi_\ep^+\neq0$.
If $\tilde{\Ga}=0$, then the zeros of the Evans function are
given by $\ga=0$ and $\ga=\pm a\ep^2$.
Upon using the inversion formula $\la=\ga^2/2+\la_b$, it is seen that 
there is an eigenvalue at $\la=0$, and no eigenvalues with positive real
part.
Thus, it is expected that the plane $\tilde{\Ga}=0$ will serve
as the critical plane for which an edge bifurcation may take place.

\begin{figure}[htb]
\epsfysize=7.5in
%%\epsffile{d:/papers/gl_front/z_0flow.ps}
\epsffile{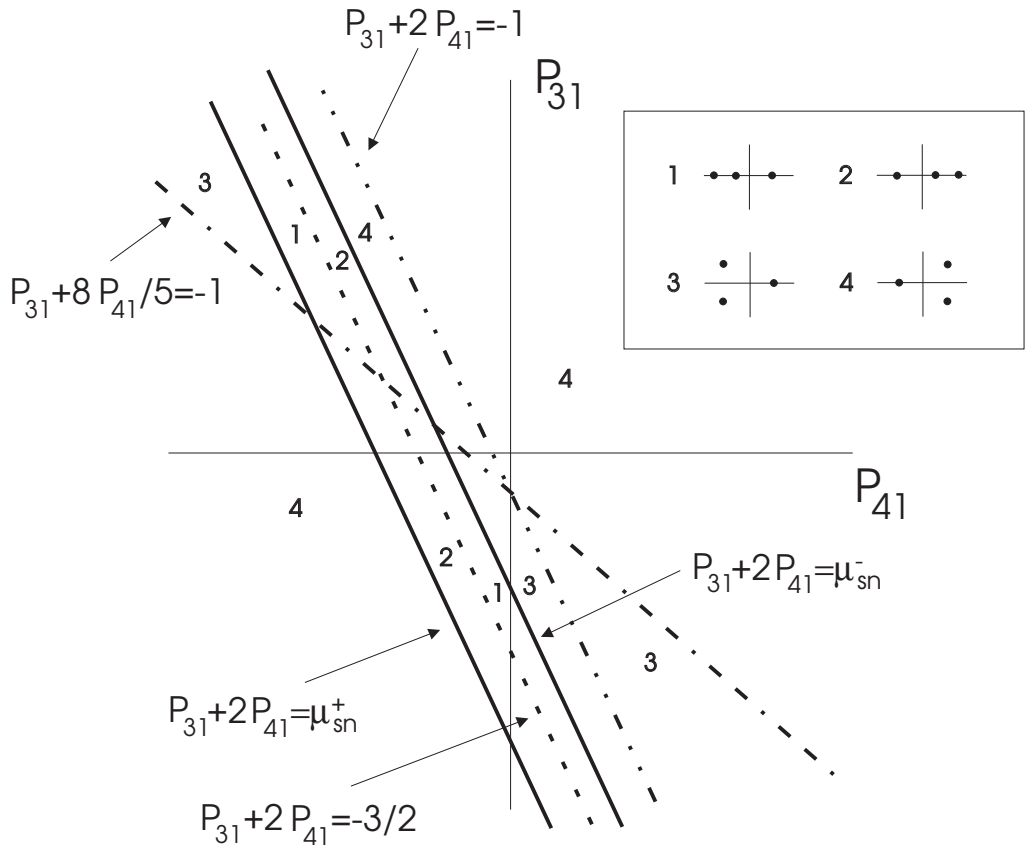}
\caption{Zeros of $E(\gamma,\epsilon)\, (d_1>0)$
     \label{fig:cgl1}}
\end{figure}

Now assume for the rest of the discussion that $\tilde{\Ga}\neq0$.
Set 
\[
\ga=\tilde{\Ga}\psi_\ep^+\ep^2 y.
\]
Solving $E(\ga,\ep)=0$ is then equivalent to solving
\[
y^3+y^2-\mu y+\mu=0,\quad \mu=\left(\frac{\Ga}{\tilde{\Ga}}\right)^2.
\]
For this equation, a saddle-node bifurcation occurs when $\mu=\al^2$.
For $0<\mu<\al^2$, there is one real negative zero, and the other
two zeros are complex with positive real part.
For $\mu>\al^2$, all of the zeros are real, and two are positive
while one is negative (see Figure \ref{fig:cgl1}).

Using the definition of the variable $y$ and the inversion formula,
it is seen that for $\Re\ga>0$,
\[
\begin{array}{lll}
\la&=&{\ds\frac12(y^2-\mu)(\tilde{\Ga}\psi_\ep^+)^2\ep^4} \\
\vspace{.1mm}\\
&=&{\ds-\frac12\frac{y^2+\mu}{y}(\tilde{\Ga}\psi_\ep^+)^2\ep^4}.
\end{array}
\]
First suppose that $\tilde{\Ga}\psi_\ep^+<0$.
To achieve a positive zero for $\ga$, one must then have $y<0$.
Since $y^2+\mu>0$, this then implies that there is a real positive
eigenvalue $\la$, so that the wave is unstable.
Now suppose that $\tilde{\Ga}\psi_\ep^+>0$.
One must then look at those roots with $\Re y>0$.
If $y$ is real, then it is clear that the resulting eigenvalues
$\la$ are negative.
If $y=y_1+iy_2$ is complex with $y_1>0$, then by checking that
\[
\Re\frac{y^2+\mu}{y}=\frac{y_1}{y_1^2+y_2^2}(y_1^2+y_2^2+\mu)>0,
\]
it is seen that the resulting complex pair of eigenvalues has negative real
part.
The picture is summarized in Figure \ref{fig:cgl2}.
Thus, the following lemma holds; Theorem \ref{thm:evals_GL} follows
from Lemma \ref{lem:evans_roots_d1=0} and this result.  

\begin{lemma}\label{lem:evans_roots_d1not0} Suppose that $d_1\neq0$,
and set $P_{j1}=d_j/d_1$. 
If
\[
(\frac32 +P_{31}+2P_{41})(1+P_{31}+\frac85P_{41})<0,
\]
then there is one positive real $O(\ep^4)$ eigenvalue
for the linearized problem, and the wave is linearly unstable.
If
\[
d_1(1+P_{31}+\frac85P_{41})>0,\quad d_1(\mu_{sn}^- +P_{31}+2P_{41})>0
\]
or 
\[
d_1(1+P_{31}+\frac85P_{41})<0,\quad d_1(\mu_{sn}^+ +P_{31}+2P_{41})<0,
\]
then there is a complex pair of $O(\ep^4)$ eigenvalues with negative
real part ($\mu_{sn}^\pm$ are defined in equation (\ref{eq:saddle-node})).
Otherwise, no eigenvalues bifurcate from the continuous spectrum.
\end{lemma}

\section{Perturbation calculations at the branch point: NLS}
\setcounter{equation}{0}

\subsection{Preliminaries}

As in the previous section, let the perturbation of the wave be written 
in the form
\[
u+iv=(r+(p+iq))e^{i\int_0^x\psi(s)\,ds}.
\]
For $\ep\neq0$, the linearized eigenvalue problem derived from 
(\ref{eq:NLS_pert}) is given up to $O(\ep^2)$ by 
\begin{equation}\label{eq:L_full_NLS}
\la\left[\begin{array}{cc}\ep d_1 & -(1-\ep^2d_1^2) \\
     1-\ep^2d_1^2 & \ep d_1  \end{array}\right]=L_0+\ep L_\ep
    +\frac12\ep^2 L_{\ep\ep},
\end{equation}
where the operators $L_0,\,L_\ep,$ and $L_{\ep\ep}$ are specified in equations
(\ref{eq:def_L0})-(\ref{eq:def_Lep2}).
As previously, the expansion for the linear operator $L$ given in
equations (\ref{eq:L_full})-(\ref{eq:def_Lep2}) yields an expansion for 
the matrix $M(\la,x)$
with $M(\la,x)\to M_\pm(\la)$ as $x\to\pm\infty$.
As in (\ref{eq:def_ag}), we set $\Ga=d_1+d_3+2d_4$ and $a=\Ga\psi_{\ep}^+$.

\begin{prop}\label{prop:branch_pt_NLS} 
The branch point of the Evans function is given by
\[
\la_b=\frac{a^2}{2(\Ga-d_1)}\ep^3.
\]
For $\la$ close to $\la_b$ the eigenvalues of $M_\pm(\la)$ which
have geometric multiplicity one and algebraic multiplicity two when
$\la=\la_b$ are given by
\[
\al_\pm^b\mp\psi_\ep^+\la(\ga)\ep\mp\ga,
\]
where
\[
\al_\pm^b=\pm a\ep^2,
\]
and 
\[
\ga=\sqrt{\la^2-2\ep(\Ga-d_1)\la+a^2\ep^4},
\]
and
\[
\la(\ga)=(\Ga-d_1)\ep+\sqrt{\ga^2+(\Ga-d_1)^2\ep^2-a^2\ep^4}.
\]
When $\la=\la_b$, the associated eigenvectors are given by
\[
\eta_\pm^b=\mp\bfu_4(0)+a\ep^2\bfu_3(0).
\]
\end{prop}

\begin{remark} To ensure that $\la_b<0$, it is necessary that
\[
\Ga-d_1=d_3+2d_4<0.
\]
This condition is consistent with \cite{burtsev:ndo97, chen:eon98, 
ikeda:tco95, ikeda:sod97}, and it will henceforth be assumed.
\end{remark}

\begin{remark} Since we are taking the principal square root, note
that up to leading order $\la(0)=\la_b$ for all $\ep\ge0$.
\end{remark}

\subsection{Calculations for $\bfy_f^\pm$}

As in Section 4.2, we use the Taylor expansions of $\bfy_f^\pm(\la,x)$,  
centered at $\la=0$, for $x$ fixed at the origin.
From (\ref{eq:yfpm}), 
\[
\partial_\ep\bfy_f^\pm(0,x)=(\psi_\ep(x)\mp\psi_\ep^+)\bfu_3(x)
  +\Phi(x)\psi_\ep'(x)\bfu_3(0),
\]
so that
\[
\partial_\la M_0(0,x)\partial_\ep\bfy_f^\pm(0,x)=
 2\Phi(x)(\psi_\ep(x)\mp\psi_\ep^+)\bfu_2(0).
\]
The expression given in equation (\ref{eq:bfyf_NLS}) implies that  
\[
M_\ep(x)\partial_\la\bfy_f^\pm(0,x)=
   2\frac{\Phi'(x)}{\Phi(x)}\psi_\ep(x)\bfu_2(0).
\]
Solving the equation
\[
(\partial_{\ep\la}^2\bfy_f^\pm)'=M_0\partial_{\ep\la}^2\bfy_f^\pm
  +M_\ep\partial_\la\bfy_f^\pm+\partial_\la M_0\partial_\ep\bfy_f^\pm
\]
by variation of parameters thus gives  
\[
\begin{array}{l}
{\ds \partial_{\ep\la}^2(\bfy_f^- -\bfy_f^+)(0,0)=2\left(
  \int_{-\infty}^{+\infty}\frac{\Phi'(x)}{\Phi(x)}\psi_\ep(x)\,dx \right.} \\
\vspace{.1mm} \\
\quad\quad\quad {\ds \left.+2\psi_\ep^+\int_{-\infty}^0\Phi(x)\Phi'(x)\,dx \right)
    \bfu_2(0)+c_1\bfu_1(0), } 
\end{array}
\]
which upon integrating yields
\begin{equation}\label{eq:bfy_la_diff}
\partial_{\ep\la}^2(\bfy_f^- -\bfy_f^+)(0,0)=
  \frac43(d_1+d_3+\frac45d_4)\bfu_2(0)+c_1\bfu_1(0).
\end{equation}
Evaluating the Taylor expansions for both $\bfy_f^- -\bfy_f^+$ and
$\partial_\la(\bfy_f^- -\bfy_f^+)$, centered at $\la=0$,
and using the fact that 
$\la_b=O(\ep^3)$ from Proposition 
\ref{prop:branch_pt_NLS} yield the following lemma (to leading order).

\begin{lemma}\label{lem:yf_NLS} The difference in the fast solutions 
satisfies
\[
\partial_\ep^4(\bfy_f^- -\bfy_f^+)(\la_b,0)=16\Ga b (\psi_\ep^+)^2\bfu_2(0)
  +c_{14}\bfu_1(0)+c_{34}\bfu_3(0),
\]
where 
\[
b=d_1+d_3+\frac45 d_4,
\]
for some constants $c_{14}$ and $c_{34}$.
Furthermore, 
\[
\partial_\ep^j(\bfy_f^- -\bfy_f^+)(\la_b,0)=c_{1j}\bfu_1(0)+c_{3j}\bfu_3(0),
  \quad j=0,\dots,3
\]
for some constants $c_{1j}$ and $c_{3j}$.
In addition
\[
\partial_{\ep\la}^2(\bfy_f^- -\bfy_f^+)(\la_b,0)=
  \frac43b\,\bfu_2(0)+c_1\bfu_1(0).
\]
\end{lemma}

\subsection{Calculations for $\bfy_s^\pm$}

The only difference in the results of Proposition \ref{prop:branch_pt_NLS}
and Proposition \ref{prop:branch_pt} arises in the
expression for the branch point $\la_b$.
Furthermore, since $|\la_b|\le O(\ep^3)$ in both cases, the
fact that it changes does not affect the calculations up to
$O(\ep^2)$.
Hence, 
the proof of Lemma \ref{lem:ys} applies here to give the following result.  

\begin{lemma}\label{lem:ys_NLS} The difference in the slow solutions 
at $\ga=0$ satisfies
\[
\partial_\ep(\bfy_s^- -\bfy_s^+)(0,0)=0,
\]
and
\[
\partial_\ep^2(\bfy_s^- -\bfy_s^+)(0,0)=
   -4(\frac12d_1+\Ga)\psi_\ep^+\,\bfu_4(0)+c_2\bfu_1(0)
\]
for some constants $c_1$ and $c_2$.
\end{lemma}

\subsection{Calculations for the Evans function}

Set
\[
\tilde{\Ga}=\frac12d_1+\Ga.
\]
In the sequel, all of the evaluations will be performed at
$(\ga,x,\ep)=(0,0,0)$, and the constants $c_i$ will be unknown
(but irrelevant).
Recall that $\partial_\ga=\partial_\la$; using this fact, along with
equation (\ref{eq:bfys_NLS}) and Lemmas \ref{lem:yf_NLS} and 
\ref{lem:ys_NLS}, we can differentiate to obtain
a perturbation expansion for the Evans function.
As in the previous section, the $\ep$-dependence of the Evans function
is being implicitly assumed.
First, we find 
\[
\begin{array}{lll}
\partial_\ep^6E(0)&=&{\ds \frac{6!}{2!4!}\partial_\ep^2(\bfy_s^- -\bfy_s^+)
  \wedge\partial_\ep^4(\bfy_f^- -\bfy_f^+)\wedge\bfy_s^+
    \wedge\bfy_f^+ } \\
\vspace{.1mm} \\
&=&{\ds \frac43 6!\,\Ga\tilde{\Ga}b(\psi_\ep^+)^3,}
\end{array}
\]
and
\[
\begin{array}{lll}
\partial_\ep^3\partial_\ga E(0)&=&{\ds \frac{3!}{1!2!}
     \partial_\ep^2(\bfy_s^- -\bfy_s^+)
  \wedge\partial_{\ep\ga}^2(\bfy_f^- -\bfy_f^+)\wedge\bfy_s^+
   \wedge\bfy_f^+}  \\
\vspace{.1mm} \\
&=&{\ds \frac83 3!\,\tilde{\Ga}b\psi_\ep^+},
\end{array}
\]
and
\[
\begin{array}{lll}
\partial_\ep\partial_\ga^2 E(0)&=&{\ds \frac{2!}{1!1!}
   \partial_\ga(\bfy_s^- -\bfy_s^+)
  \wedge\partial_{\ep\ga}^2(\bfy_f^- -\bfy_f^+)\wedge\bfy_s^+
   \wedge\bfy_f^+} \\
\vspace{.1mm} \\
&=&{\ds -\frac{8}{3} 2!\,b. }
\end{array}
\]
In addition, recall equation (\ref{eq:E_ggg_NLS}), which states that
\[
\partial_\ga^3 E(0)=-24.
\]
All lower derivatives of $E$ are zero, so based on
the above expansions, the Evans function can
be written as
\begin{equation}\label{eq:evans_expand_a<0_NLS}
\begin{array}{lll}
E(\ga,\ep)&=&{\ds -4(\ga^3+\frac23b\ep\ga^2-
    \frac23\tilde{\Ga}b\psi_\ep^+\ep^3\ga
  -\frac13\Ga\tilde{\Ga}b(\psi_\ep^+)^3\ep^6)} \\
\vspace{.1mm} \\
&=&{\ds -4(\ga+\frac23b\ep)(\ga^2-\tilde{\Ga}\psi_\ep^+\ep^2\ga
   -\frac12\Ga\tilde{\Ga}(\psi_\ep^+)^3\ep^5)}.
\end{array}
\end{equation}

To leading order, the roots are for the Evans function are thus
\begin{equation}\label{eq:evans_roots_NLS}
\ga_1=-\frac23b\ep,\,\, \ga_2=\tilde{\Ga}\psi_\ep^+\ep^2,\,\,
   \ga_3=-\frac12\Ga(\psi_\ep^+)^2\ep^3.
\end{equation}
These can correspond to true eigenvalues only if
$\Re\ga>0$.
First suppose that $b<0$, so that $\ga_1>0$.
From the transformation given in Proposition 
\ref{prop:branch_pt_NLS}, i.e.,
\[
\la(\ga)=(\Ga-d_1)\ep+\sqrt{\ga^2+(\Ga-d_1)^2\ep^2-a^2\ep^4},
\]
we find, to leading order, the positive eigenvalue
\begin{equation}\label{eq:eval1_NLS}
\la_1=-(\Ga-d_1)\left(\sqrt{1+\frac{4b^2}{9(\Ga-d_1)^2}}-1\right)\,\ep.
\end{equation}
Now suppose that $\tilde{\Ga}\psi_\ep^+>0$, so that $\ga_2>0$, and 
set $\ga_2^2-a^2\ep^4=\tilde{\ga}\ep^4$,
where
\[
\tilde{\ga}=d_1(\psi_\ep^+)^2(\frac54d_1+d_3+2d_4).
\]
One obtains, to leading order, the second eigenvalue
\begin{equation}\label{eq:eval2_NLS}
\la_2=-\frac{\tilde{\ga}}{2(\Ga-d_1)}\ep^3, 
\end{equation}
which is only positive if $\tilde{\ga}>0$.
Finally, independent of its sign, $\ga_3$ is of too high an order to 
correspond to 
a positive eigenvalue $\la$; hence, it can be ignored.
The following lemma has now been proved; this also yields Theorem 
\ref{thm:evals_NLS}.

\begin{lemma}\label{lem:evans_roots_NLS} Let $d_3+2d_4<0$.
Suppose that $d_1>0$,
and set $P_{j1}=d_j/d_1$.
If 
\[
P_{31}<-\frac45P_{41}-1,
\]
then there is a positive $O(\ep)$ real eigenvalue given, to leading order, by 
equation
(\ref{eq:eval1_NLS}).
Furthermore, if
\[
P_{31}>-\frac85P_{41}-1,\quad P_{31}>-2P_{41}-\frac54,
\]
then there is a positive $O(\ep^3)$ real eigenvalue which is given, to 
leading order, by
equation (\ref{eq:eval2_NLS}).
Otherwise, the wave is linearly stable, as no other eigenvalues 
bifurcate from the continuous spectrum (see Figure \ref{fig:nls}).
If $d_1=0$, then the wave is linearly stable if $5d_3+4d_4<0$; otherwise,
there is an $O(\ep)$ eigenvalue which is given by equation 
(\ref{eq:eval1_NLS}).
\end{lemma}

\subsection{Comparison with adiabatic approach}

There have been many recent efforts to determine the stability of the
dark soliton for the perturbed NLS by using an adiabatic approach
(\cite{burtsev:ndo97, chen:eon98, ikeda:tco95, ikeda:sod97, lega:ths97}).
Following Lega \etal \cite{lega:ths97}, write the solution to the
perturbed NLS as 
\[
\begin{array}{l}
\phi=(\kappa R\Phi(\kappa\xi)+\ep\phi_1+\ep^2\phi_2+\cdots)\,\times \\
\vspace{.1mm} \\
\quad\quad\quad   \exp[i(qx-\Omega t+qx_0+\theta_0)]
   \exp[i\ep\int_0^{\kappa\xi}\psi_\ep(s)\,ds],
\end{array}
\]
where
\[
\xi=x-ct+x_0,\,\,q=k\kappa-c,\,\,\Omega=-\frac12q^2-(\kappa R)^2,\,\,
  R^2=1+k^2.
\]
Following the procedure outlined in Appendix C of \cite{lega:ths97},
and using the requirement that $d_2+d_3+d_4=O(\ep^2)$ for
the dark soliton to persist as a regular perturbation, one finds
that for the time scale $T=\ep t$,
\[
\begin{array}{lll}
k_T&=&{\ds \frac23\kappa[d_1c-(d_1+d_3)k\kappa-\frac65d_4k\kappa^3
   (1+\frac53k^2)](1+k^2)} \\
\vspace{.1mm} \\
\kappa_T&=&{\ds [d_3(\kappa^2-1)+d_4(\kappa^4-1)+
           (d_3+2d_4)k^2\kappa^2} \\
\vspace{.1mm} \\
&&\quad\quad {\ds +d_4k^4\kappa^4-\frac12d_1q^2]\kappa
           -\frac{k}{1+k^2}\kappa k_T}.
\end{array}
\]

A linear stability analysis of the critical point $(k,\kappa,c)=(0,1,0)$
yields the eigenvalues
\[
\la_1=2(d_3+2d_4),\,\,\la_2=-\frac23(d_1+d_3+\frac65d_4).
\]
Thus, with this approach the wave is claimed to be stable if both $d_3+2d_4<0$ 
and $d_1+d_3+6d_4/5>0$ hold.
If $d_4=0$, then this analysis is consistent with the result of
Lemma \ref{lem:evans_roots_NLS} in that it correctly determines the
stability of the wave up to $O(\ep)$.
However, and this is not surprising, it does not predict the existence
of the $O(\ep^3)$ instability.
If $d_4\neq0$, then the analysis is consistent with what was
found via the adiabatic approach
in \cite{burtsev:ndo97, chen:eon98, ikeda:tco95, ikeda:sod97};
however, these all contradict the results presented in this 
paper, even at the $O(\ep)$ level.
This contradiction implies that the original ansatz for the
slow-time variation of the wave in the adiabatic approach is incorrect.
In some way the parameter $d_4$ has the same effect on
the stability analysis for the perturbed wave as it has on  
the solution structure for the steady-state problem,
i.e., it breaks some kind of ``hidden symmetry'' (see
Doelman \cite{doelman:bth96}).
As mentioned in the Introduction, this would be an interesting topic
for future study.

\ack
We thank Bj\"orn 
Sandstede for several stimulating conversations.
We also thank Alejandro Aceves for an illuminating discussion
on the use of the adiabatic method for stability analysis.
Finally, we thank Evans Compton for a great day at the beach.
The research of T. Kapitula is partially supported under 
NSF grant DMS-9803408, and the research of J. Rubin is
partially supported under NSF grant DMS-9804447.

\vspace{5mm}

%\input{concl}

%\appendix
%\renewcommand{\theequation}{\Alph{section}.\arabic{equation}}
%\newpage
%\input{app1}
%\input{app2}

%%%%%%%%%%%%%%%%%%%%%%%%%%%%%%%%%%%%%%%%%%%%%%%%%%%%%%%%%%%%%%%%%%%%%%%%
\bibliography{papers}
%\bibliography{../../papers}
%\bibliography{d:/papers/papers}
\bibliographystyle{plain}
%%%%%%%%%%%%%%%%%%%%%%%%%%%%%%%%%%%%%%%%%%%%%

%%%%%%%%%%%%%%%%%%%%%%%%%%%%%%%%%%%%%%%%%%%55
\newpage

\newpage

\newpage

\newpage

%%%%%%%%%%%%%%%%%%%%%%%%%%%%%%%%%%%%%%%%%%%%%

\end{document}